\documentclass[a4paper,10pt,         
pointlessnumbers,      
twoside,
DIV=calc,               
BCOR15mm               
]{article}
\usepackage{geometry}
\usepackage[T1]{fontenc}        

\usepackage[utf8x]{inputenc}    
\usepackage{ucs}                
\usepackage{amsmath}            
\usepackage{amsfonts}
\usepackage{amssymb}
\usepackage{amsthm} 
\usepackage{thmtools}
\usepackage[authoryear]{natbib}
\usepackage{ifthen}
\usepackage{color}
\usepackage{xcolor}	
\usepackage{colortbl}
\usepackage{graphicx} 
\usepackage{textcomp}            
\usepackage{palatino}            
\usepackage{hhline}

\usepackage{varioref}            
\usepackage{url}                
\usepackage{cleveref}             
\usepackage{fancyhdr}
\usepackage{float}
\usepackage{multirow}
\usepackage{rotating}
\usepackage{subfig}
\usepackage{titlesec}
\usepackage{setspace}
\usepackage{tabularx}
\usepackage{booktabs}
\definecolor{tugreen}{HTML}{84b819}
\definecolor{pgrey}{rgb}{242, 242, 242}
\definecolor{ashgrey}{rgb}{0.7, 0.75, 0.71}
 	\definecolor{gainsboro}{rgb}{0.86, 0.86, 0.86}
 	\definecolor{lightyellow}{rgb}{1.0, 1.0, 0.88}
 	\definecolor{pastelgray}{rgb}{0.81, 0.81, 0.77}
 	\definecolor{timberwolf}{rgb}{0.86, 0.84, 0.82}
 	\definecolor{beige}{rgb}{0.96, 0.96, 0.86}
\titleformat{\chapter}{\fontfamily{qzc}\selectfont \Huge \color{tugreen}}{\thechapter}{1ex}{}
\titlespacing{\chapter}{0pt}{-2ex}{*1.5}
\usepackage{xr} 
\usepackage{pdfpages}
\usepackage{xcite}

\newcommand{\R}{{\mathbb{R}}}         
\newcommand{\E}{{\mathbb{E}}}
 \newcommand{\N}{{\mathbb{N}}}         

\newcommand{\bay}{\begin{array}}
\newcommand{\eay}{\end{array}}

\newcommand{\bqa}{\begin{eqnarray*}}
\newcommand{\eqa}{\end{eqnarray*}}

\newcommand{\bee}{\begin{eqnarray*}}
\newcommand{\eee}{\end{eqnarray*}}

\newcommand{\bea}{\begin{eqnarray*}}
\newcommand{\eea}{\end{eqnarray*}}

\newcommand{\bqan}{\begin{eqnarray}}
\newcommand{\eqan}{\end{eqnarray}}

\newcommand{\be}{\begin{eqnarray}}
\newcommand{\ee}{\end{eqnarray}}

\newcommand{\bit}{\begin{itemize}}
\newcommand{\eit}{\end{itemize}}

\newcommand{\ben}{\begin{enumerate}}
\newcommand{\een}{\end{enumerate}}

\newcommand{\beq}{\begin{equation}}
\newcommand{\eeq}{\end{equation}}

\newcommand{\bdes}{\begin{description}}
\newcommand{\edes}{\end{description}}

\newcommand{\btb}{\begin{tabular}}
\newcommand{\etb}{\end{tabular}}

\newcommand{\bcen}{\begin{center}}
\newcommand{\ecen}{\end{center}}

\newcommand{\bmp}{\begin{minipage}}
\newcommand{\emp}{\end{minipage}}

\newcommand{\vepsilon}{\vep}

\newcommand{\Cov}{\operatorname{{\it Cov}}}

\newcommand{\tr}{\operatorname{tr}}

\newcommand{\diag}{\operatorname{\it diag}}

\newcommand{\va}{\boldsymbol{a}}
\newcommand{\vb}{\boldsymbol{b}}
\newcommand{\vc}{\boldsymbol{c}}

\newcommand{\ve}{\boldsymbol{e}}
\newcommand{\vf}{\boldsymbol{f}}
\newcommand{\vg}{\boldsymbol{g}}
\newcommand{\vh}{\boldsymbol{h}}

\newcommand{\vr}{\boldsymbol{r}}

\newcommand{\vv}{\boldsymbol{v}}

\newcommand{\vx}{\boldsymbol{x}}

\newcommand{\vA}{\boldsymbol{A}}
\newcommand{\vB}{\boldsymbol{B}}
\newcommand{\vC}{\boldsymbol{C}}
\newcommand{\vD}{\boldsymbol{D}}
\newcommand{\vE}{\boldsymbol{E}}

\newcommand{\vI}{\boldsymbol{I}}
\newcommand{\vJ}{\boldsymbol{J}}

\newcommand{\vL}{\boldsymbol{L}}
\newcommand{\vM}{\boldsymbol{M}}
\newcommand{\vN}{\boldsymbol{N}}

\newcommand{\vP}{\boldsymbol{P}}

\newcommand{\vR}{\boldsymbol{R}}

\newcommand{\vT}{\boldsymbol{T}}

\newcommand{\vV}{\boldsymbol{V}}

\newcommand{\vX}{\boldsymbol{X}}
\newcommand{\vY}{\boldsymbol{Y}}
\newcommand{\vZ}{\boldsymbol{Z}}

\newcommand{\vep}{\boldsymbol{\epsilon}}

\newcommand{\vmu}{\boldsymbol{\mu}}

\newcommand{\vSigma}{\boldsymbol{\Sigma}}

\newcommand{\vtheta}{\boldsymbol{\theta}}
\newcommand{\vTheta}{\boldsymbol{\Theta}}
\newcommand{\vzeta}{\boldsymbol{\zeta}}

\newcommand{\veins}{{\bf 1}}
\newcommand{\vnull}{{\bf 0}}

\newcommand{\vUpsilon}{\boldsymbol{\Upsilon}}

\newcommand{\ind}{1\hspace{-0.7ex}1}

\newcommand{\Cdot}{\cdot}

\DeclareMathOperator{\vech}{vech}
\DeclareMathOperator{\dvech}{dvech}

\newtheoremstyle{Test1}
  {2 \baselineskip}
  {1.5 \baselineskip}
  {\itshape}
  {-0.0ex}
  {\fontfamily{ppl}\fontseries{l}\fontshape{n}}
  {:}
  {\newline}
   {}

\theoremstyle{Test1}
\newtheorem{Sa}{Theorem}[section]
\newtheorem{exmp}{Example}[section]
\newtheorem{theorem}{Theorem}[section]
\newtheorem{re}[Sa]{Remark}

\newtheorem{Ko}[Sa]{Corollary}

\newcommand{\lan}{ \scriptstyle \mathcal{O}\textstyle}

\newcolumntype{x}[1]{!{\centering\arraybackslash\vrule width #1}}
\makeatletter
\renewenvironment{proof}[1][\proofname]{\par
  \pushQED{\qed}%
  \fontfamily{ppl}\fontseries{m}\fontshape{it} \topsep6\p@\@plus6\p@\relax
  \trivlist
  \item[\hskip\labelsep
        \bfseries
    #1\@addpunct{:}]\ignorespaces
}{%
  \popQED\endtrivlist\@endpefalse
}
\makeatother

\usepackage[skip=2pt plus1pt, indent=10pt]{parskip}
\begin{document}


\title{\Large \bf Testing for patterns and structures in covariance and correlation matrices}
\author{Paavo Sattler$^{1}$ and Dennis Dobler$^{1,2}$ 
\\[0.4ex] }

\vspace{-8ex}
  \date{}
\maketitle \vspace*{0.25ex}
\begin{center}\noindent${}^{1}$ TU Dortmund University, Department of Statistics, Germany\\
\noindent${}^{2}$ Research Center Trustworthy Data Science and Security, University Alliance Ruhr, Germany
\\\mbox{ }\hspace{1 ex}email: paavo.sattler@tu-dortmund.de  \\
  \end{center}

\begin{abstract}
\noindent
Covariance matrices of random vectors contain information that is crucial for modelling. Specific structures and patterns of the covariances (or correlations) may be used to justify parametric models, e.g., autoregressive models. Until now, there have been only a few approaches for testing such covariance structures and most of them can only be used for one particular structure. In the present paper, we propose a systematic and unified testing procedure working among others for the large class of linear covariance structures.
Our approach requires only weak distributional assumptions. It covers common structures such as diagonal matrices, Toeplitz matrices and compound symmetry, as well as the more involved autoregressive matrices. We exemplify the approach for all these structures.
We prove the correctness of these tests for large sample sizes and use bootstrap techniques for a better small-sample approximation. Moreover, the proposed tests invite adaptations to other covariance patterns by choosing the hypothesis matrix appropriately.
 With the help of a simulation study, we also assess the small sample properties of the tests. 
 Finally, we illustrate the procedure in an application to a real data set.

\end{abstract}


\vfill
\vfill

\section{Motivation and Introduction}\label{int}

Tests about variances have many possible applications, not only as pretests but also as stand-alone tests for situations where effects on the variation (instead of the location) are investigated. In a multivariate setting, dependency structures in terms of covariances play an important role next to the variances on the diagonal of the covariance matrix. 
In this context, checking whether a covariance matrix has a special structure like compound symmetry or sphericity is of great interest.

Because of this topic's relevance, several more or less detailed approaches exist. For instance, under normality, \cite{zhong2017} considered general hypotheses for high-dimensional data, while 
\cite{votaw1948} and \cite{winer1991} provided test procedures for a compound symmetry structure of the covariance matrix. In a special model under normality, \cite{mckeown1996} considered the combined hypothesis of the equality of covariance matrices and a first-order autoregressive structure in multi-sample comparisons.
For non-high-dimensional data, \cite{gupta2006} developed a test for sphericity, which works under some conditions on the characteristic function.
Finally, the methods of \cite{wakaki1990} allow for testing a larger class of structures with fewer distributional restrictions. Unfortunately, their procedure is quite complex since many parameters need to be calculated or estimated. Together with the fact that the procedure was not illustrated by means of a concrete test application,  this approach is challenging to use in practice \citep{yuan2005,herzog2007}.

Thus, we aim to develop a new test which allows the user to test a wide range of possible structures with fewer distributional conditions, and at the same time, it will be intuitive and appealing.
To this end, we will use a quite general test for hypotheses regarding covariances as a starting point \citep{sattler2022}. That paper introduced a nonparametric model with few distributional assumptions, some hypotheses, and suitable hypothesis matrices. Due to the great generality of the considered model, extensions to various other hypotheses are readily developed. 
For example, this concerns particular patterns in the covariance matrix or particular structures of covariance matrices. Since there exist different notions for patterns (see e.g. \cite{graybill1983} and \cite{kollo2005}), we will focus on the latter; we will explain the difference between patterns and structures later on.
We will show that our test approach can be used in a wide and commonly used class of structures and how the corresponding hypothesis can be formulated.

Next to covariance matrix structures, correlation matrix structures are also often of interest. As in \cite{sattler2022}, we will exploit a link between tests regarding correlations and covariances \citep{sattler2023correlation} to expand the approach to cover tests about correlation matrices as well. Finally, also combined tests for multiple structures are possible, and the according procedure is introduced.

 This paper is organized as follows: in \Cref{Modell and Statistics}, we introduce the model and recapitulate the results of \cite{sattler2022}, which are used to develop tests for structures of covariance matrices. Afterwards, in  \Cref{Linear Covariance Structure Model}, we present a test procedure which can be used for a large class of covariance matrix structures together with the required hypothesis matrices for their most important representatives. Afterwards, in \Cref{section:Extended linear covariance structure model}, we also extend this to a larger class which, for instance, includes the autoregressive structure. Next, testable structures of correlation matrices are treated in \Cref{Structures of correlation matrices}. Since many structures result from an overlap of other, simpler structures, we outline in \Cref{Combined} a combined test procedure for the more complex structures.
In \Cref{Simulations}, we present simulation results to assess the type-I error control and the power of the test.
Finally, we illustrate the methodology with an application to EEG data in Section~\ref{Illustrative Data Analysis}.
 We will conclude with a discussion in Section~\ref{conclude}.
 All proofs and additional simulation results are offered in supplementary material available online.

\section{Model and Statistics} \label{Modell and Statistics}

We consider a general semiparametric model given by independent $d$-dimensional random vectors \bqan\label{model}
\vX_{k}&= \vmu + \vep_{j}, 
\eqan
where $k=1, \dots, N$ refers to the individual with the $d$-dimensional measurement outcome $\vX_k$.
In this setting, $\E(\vX_{1})=\vmu = (\mu_{1}, \dots, \mu_{d})^\top \in \mathbb{R}^d$, and
the errors $\vep_{1},\dots,\vep_{N}$ are assumed to be i.i.d.\ and centered, $\E(\vep_{1}) = \vnull_d$,   with a positive semidefinite covariance matrix $\Cov(\vep_1)=\vV  \in \mathbb{R}^{d \times d}$. Moreover, we require 
the fourth moment to be finite, $\E(||\vep_{1}||^4) < \infty$, where $ || \cdot ||$ denotes the Euclidean norm. 
No further distributional assumptions for the error terms are required.
Due to the symmetry of covariance matrices, 
it suffices to analyze their structures based on the upper triangular components.
This reduces the essential components of 
$\vV =(v_{rs})_{r,s}^d$ to a $p:=d(d+1)/2$-dimensional subset.

For simplicity, one may focus on the vectorization of the upper triangular covariance matrix, say ``the vectorized half''   $\vech(\vV):= (v_{11},v_{12},\dots,v_{1d},v_{22},\dots,v_{2d},\dots,v_{dd})^\top$; cf.\ \cite{sattler2022}.
Under the assumed moment conditions, it is well-known that the empirical covariance matrix, $\widehat \vV = \sum_{j=1}^N (\vX_j - \overline \vX)(\vX_j - \overline \vX)^\top/(N-1)$ with $\overline \vX = \sum_{j=1}^N \vX_j/N$, follows a central limit theorem.
In the context of the vectorized half, this convergence translates to
\begin{equation}\label{Nv1}
\sqrt{N} (\vech(\widehat \vV) -\vech(\vV))\stackrel{\mathcal {D}}{\longrightarrow} {\mathcal{N}_{p}\left(\vnull_{p},\vSigma \right)} \quad \text{as \ } N \to \infty,\end{equation}
while \cite{sattler2022} also developed a consistent estimator $\widehat\vSigma  \ \stackrel{\mathcal{P}}{\to} \ \vSigma=\Cov(\vech(\vepsilon_{1}\vepsilon_{1}^\top)) \in \mathbb{R}^{p\times p} $.
Here $\stackrel{\mathcal{D}}{\to}$ denotes convergence in distribution and $\stackrel{\mathcal{P}}{\to}$ convergence in probability, both as $N\to \infty$.\\
However, for analyzing most covariance and correlation matrix structures, it is more convenient to vectorize along the diagonals and then along secondary diagonals, instead of a line-by-line vectorization. 
For this, we introduce 
$$\vv=\dvech(\vV):=(v_{11},v_{22},...,v_{dd},v_{12},....,v_{(d-1)d},....,v_{1d})^\top$$
and we define $\widehat \vv=\dvech(\widehat \vV)$. Since $\vech$ and $\dvech$ differ only in the order of the elements, there exists a permutation matrix $\vA\in\R^{p\times p}$ with $ \dvech(\vB)=\vA\vech(\vB)$ for any matrix $\vB\in \R^{d\times d}$. Therefore, all results from \cite{sattler2022} implicitly also hold for $\dvech$ instead of $\vech$, resulting in the following theorem.

\begin{theorem}[\cite{sattler2022}]\label{TheoremAlt}
   As $N\to \infty$, we have convergence in distribution
\begin{equation}\sqrt{N} (\widehat \vv -\vv)\stackrel{\mathcal {D}}{\longrightarrow} {\mathcal{N}_{p}\left(\vnull_{p},\vSigma_{\dvech} \right)}\end{equation}
with $\vSigma_{\dvech}=\Cov(\dvech(\vepsilon_{1}\vepsilon_{1}^\top)) \in \mathbb{R}^{p\times p}$.
   
\end{theorem}

A consistent estimator for the unknown covariance matrix $\vSigma_{\dvech}$ is given by $\widehat \vSigma_{\dvech}=\vA\widehat \vSigma\vA^\top$, where the concrete form of $\vA$ can be found in the supplementary material.

This central limit theorem paves the way for testing hypotheses about covariance matrices,
 $\mathcal H_0^{\vv}: \vC \vv = \vzeta\in \R^{m}$, with a so-called hypothesis matrix $\vC\in\R^{m\times p}$.

 A test may be based on a so-called  ANOVA-type (test) statistic (ATS) for $\vv$, which is defined by
 \[ATS_{\vv}(\widehat{\vSigma}_{\dvech})= N\left[  \vC\widehat \vv - \vzeta\right]^\top\left[  \vC\widehat \vv - 
\vzeta\right]/\tr\left(\vC\widehat{\vSigma}_{\dvech}\vC^\top\right).\]

Together with appropriate  $(1-\alpha)$-quantiles as critical values, this leads to asymptotically correct and consistent tests.
In particular, the tests  $\varphi_{ATS}:=\ind\{ ATS_{\vv}(\widehat \vSigma_{\dvech}) >q_{1-\alpha}^{MC}\}$ and $\varphi_{ATS}^*=\ind\{ ATS_{\vv}(\widehat \vSigma_{\dvech}) >q_{1-\alpha}^*\}$ showed good performance in \cite{sattler2022}, where
 $q_{1-\alpha}^{MC}$ is found through a Monte Carlo procedure  and $q_{1-\alpha}^*$ is based on a parametric bootstrap.

In the present paper, we will focus on hypothesis tests that are based on such ATS, where the hypothesis matrix $\vC$ depends on the structure that is to be tested. But even if a structure cannot be directly formulated by  $\mathcal H_0^{\vv}: \vC \vv = \vzeta\in \R^{m}$, it can be tested based on the above central limit theorem, if there exists an appropriate function in $\vv$ based on which the null hypothesis can be formulated.

This \cref{Kof} actually includes two bootstrap approaches and will considerably expand the usage of the test, as it will be presented in the following sections.

\section{Linear Covariance Structure Model}\label{Linear Covariance Structure Model} 
From now on, we will consider one of the most important models for covariance matrices: the so-called linear covariance structure model, comparable to  \cite{anderson1973}, \cite{szatrowski1980} or \cite{zwiernik2017}.

For $q<p$ this model is given through
\begin{align}
 \label{eq:lcsmodel}
  \mathcal{V}=\{\vV(\vtheta)\in \R^{d\times d}:\vV(\vtheta)=\vV_0+\theta_1 \vV_1+...+\theta_q \vV_q, \ \vtheta = ( \theta_1, \dots, \theta_q)^\top\in \R^q\} \cap \mathcal{COV}_{d \times d}
\end{align}
where $\vV_{{0}},...,\vV_q$ are known linearly independent symmetric matrices with $\vV_1,...,\vV_q\neq\vnull_{d\times d}$ and $\mathcal{COV}_{d \times d} \subset \R^{d \times d}$ is the cone of covariance matrices, i.e., symmetric positive semi-definite matrices.

For $q=p$ and then necessarily $\vV_0 = \boldsymbol{0}_{d \times d}$  we would have $\mathcal{V}= \mathcal{COV}_{d \times d}$ and therefore no specific structure, which leads to a trivial hypothesis.
Effectively, the intersection with this cone results in a restriction of the parameter space to $\vTheta:=\{\vtheta \in \R^q:\vV(\vtheta)\geq 0\}$.
The parameter space $\vTheta$ usually has infinite cardinality. 
Our aim is to reformulate the null hypothesis $\mathcal H_0^{\vv} : \vC \vv = \vzeta $  in terms of $\vv$ being a member of the upper triangle vectorization of $\mathcal{V}$.

Of note, our approach below will not require the estimation of $\vtheta$.
Instead, we will develop tests which allow for checking the structures of $\vv$ irrespective of the specific values of $\vtheta$.
We would also like to stress that, since no further restrictions on $\vTheta$ are taken, no special relations between the components of $\vtheta$ are allowed, for example the equality of all components.
Such relationships between the components of $\vtheta$ would be required for more complex covariance matrix structures, e.g., the class of autoregressive models.
We will investigate these matters in \Cref{section:Extended linear covariance structure model}.

The literature often neglects the matrix $\vV_0$ in model formulations.
However, this matrix allows for particular structures, for example, the specification that the diagonal elements of a covariance matrix equal~1. But even without the inclusion of $\vV_0$ in the definition of $\mathcal V$, the model would contain many of the most frequently used covariance structures like diagonality or compound symmetry; 
we will investigate particular structures in the subsections to come.
In addition, the choice of $\vV_0=\vnull_{d\times d}$ would result in $\mathcal{V}$ being a conical subset of a vector subspace of $\R^{d \times d}$.
In that case, $\vTheta \subset \R^q$ defines a conical linear subspace since, for $\vtheta_1, \vtheta_2 \in \vTheta$ and $\lambda, \mu \geq 0$, $\lambda\vV(\vtheta_1) + \mu \vV(\vtheta_2) \in \mathcal V$, i.e., $\lambda \vtheta_1 + \mu \vtheta_2 \in \vTheta$.

For an arbitrary $\vV_0$, however, $\mathcal{V}$ is only an affine conical subspace. This affine subspace is only closed under affine combinations, e.g., $\lambda \vV(\vtheta_1) + \mu \vV(\vtheta_2) - (\lambda+\mu-1) \vV_0 \in \mathcal V$, but not under regular summation. This is reasonable since, for instance, the sum of two matrices with only ones on the diagonal has diagonal elements differing from one.

Since the relationships between different covariance structure model components are only additive, covariance matrices from model~\eqref{eq:lcsmodel} can be tested with our approach by using the suitable hypothesis matrix.
This is summarized in the following theorem.


\begin{theorem}\label{TheoremFormulation}
 Let $\mathcal V$ be a covariance matrix structure defined in~\eqref{eq:lcsmodel}.
 Then there exists a matrix $\vC\in\R^{(p-q)\times p}$ and a vector $\vzeta=\vC\dvech(\vV_0)\in\R^{p-q}$  such that for $\vV\in\mathcal{COV}_{d \times d}$ it holds $\vV \in \mathcal V$ if and only if 
 $\vC \cdot \textnormal{dvech}(\vV) = \vzeta$.
 In particular, the null hypothesis of having this structure can be identified with the null hypothesis  
 $\mathcal H_0^{\vv}: \vC \vv = \vzeta$, where $\vv=\textnormal{dvech}(\vV)$.
 
\end{theorem}

Hereby, for the same structure $\mathcal V$, a variety of choices for the hypothesis matrix $\vC$ is possible. The proof of \Cref{TheoremFormulation} motivates a constructive algorithm for deriving a possible matrix $\vC$, which can be exercised in every statistical programming language. 
For the R-computing environment (\cite{R}), we offer a software solution at \\ \url{https://github.com/PSattlerStat/StructureTestCovCorrMatrix}.

However, this construction method often leads to hypothesis matrices which are neither entirely intuitive nor preferable regarding the power of the resulting test. 

\begin{re}
 There exist different definitions of patterns, while we follow \cite{kollo2005}, where a patterned matrix means that only a part of the elements of a matrix is considered. So in our context a structure involves all components 
 of the matrix, while patterns only target a part of the matrix. Common examples for patterns therefore would be diagonal matrices or tridiagonal matrices. However, since the linear covariance structure and therefore also  \Cref{TheoremFormulation} work without distinction of both, the difference is only a linguistic one. This also holds for the extended linear covariance structure model introduced below, as well as the corresponding versions for correlation matrices.\\
\end{re}

\begin{re}
The literature often mentions the applicability of an approach for testing specific structures or patterns of covariance/correlation matrices without precisely explaining how the hypothesis can be represented in technical terms, for example, with the help of concrete hypothesis matrices.
So, e.g., \cite{steiger1980} investigated the hypothesis whether a correlation matrix has the structure of a Toeplitz matrix. 
Unfortunately, neither the representation of this hypothesis in the underlying model nor the suitable hypothesis matrix was mentioned.
However, this would have enhanced the readability and helped to avoid extensive definitions. At the same time, it has drawbacks, for instance, practitioners might struggle with implementing such a test procedure. Therefore, the following sections aim to provide comprehensible and suitable hypothesis matrices for the most common covariance structures. This can be seen as an example of how the hypothesis matrices can be chosen in \Cref{TheoremFormulation} and at the same time provide ideas for other related structures.
\end{re}

All of the following concrete hypothesis matrices $\vC$ are symmetric and even projection matrices. In general, however, neither is required.
In the following subsections,  it is possible to use $\vzeta=\vnull_m$ in the null hypotheses $\mathcal H_0^{\vv}: \vC \vv = \vzeta$. 
As a consequence, a unique projection matrix exists for these testing problems. To reduce the computation time (e.g., in Monte-Carlo simulations), we recommend adjusting these hypothesis matrices; see  \cite{sattler2023hypothesis} and \cite{sattler2025} for details.
Also for $\vzeta\neq \vnull_m$ or equivalently  $\vV_0\neq \vnull_{d\times d}$  such adjustments are useful, but a unique projection matrix might not exist. 

In such cases, one should be alert of the influence of the choice of the hypothesis matrices.

\subsection{Diagonality}
\label{ssec:diag}
The diagonality of a covariance matrix is by far the simplest pattern of a covariance matrix.
Nevertheless, it is still the topic of many articles, also in the last ten years; see, e.g., \cite{lan2015}, \cite{xu2017}, and \cite{touloumis2021}. Since it entails that all components are uncorrelated, it allows many conclusions about the underlying model. Hence, all non-diagonal elements are required to be zero, while no conditions on the diagonal elements are imposed.
 
The hypothesis of diagonality can be represented by $\mathcal{H}_0^{\vv}(\text{D}):\vC_D\vv=\vnull_{p}$ using the hypothesis matrix 

\[ \vC_D=\vnull_{d\times d}\oplus \vI_{p_u}, \]
where $\vnull_{d\times d} \in \R^{d \times d}$ is the zero matrix, 
$\vI_{p_u} \in \R^{p_u \times p_u}$ is the identity matrix, $p_u=p-d$, and $\oplus$ denotes the direct sum.
To remind the reader, the first $d$ components of $\vv$ contain the diagonal entries of $\vV$, followed by the off-diagonal entries.
Consequently, the part $\vnull_{d\times d}$ of the hypothesis matrix makes clear that the specific values on the diagonal do not play a role in testing the present null hypothesis.
On the other hand, the part $\vI_{p_u}$ ensures that each off-diagonal entry is tested to be zero.

\subsection{Sphericity}
Sphericity is a necessary assumption in many repeated measurement approaches, e.g., ANOVA.
Existing test procedures for sphericity can be found in the literature, for example, in \cite{gupta2006}.
The sphericity of a covariance matrix is satisfied when it equals a re-scaled identity matrix.
Therefore, it is a special case of a diagonal matrix. For this reason, we use a hypothesis matrix similar to $\vC_D$ of the previous Section~\ref{ssec:diag}, but with a matrix to check that the diagonal elements are equal. 
To this end, we replace $\vnull_{d\times d}$ in the definition of the previous section's hypothesis matrix with $\vP_d=\vI_d-\veins_d\veins_d^\top/d$, where $\veins_d \in \R^d$ is the vector with ones in each entry. The resulting hypothesis matrix is
\[\vC_{S}=\vP_{d}\oplus \vI_{p_u},\] which allows to express the hypothesis as
$\mathcal H_0^{\vv}(\text{S}):\vC_{S}\vv=\vnull_{p}$.

\subsection{Compound Symmetry}

The widespread covariance matrix structure of compound symmetry, mainly known from split-plot designs, is characterized by two conditions: the equality of
all diagonal elements and the equality of all non-diagonal entries.
As a consequence, the
appropriate hypothesis matrix is also composed of two parts through
\[\vC_{CS}=\vP_d\oplus \vP_{p_u}.\]
With this matrix, we can formulate the hypothesis that the covariance matrix is a compound symmetry matrix through $\mathcal H_0^{\vv}(\text{CS}):\vC_{CS}\vv=\vnull_{p}$. Moreover, the sphericity of a matrix can be seen as a special case of compound symmetry.

\subsection{Toeplitz}\label{Toeplitz}
Toeplitz matrices are defined by equal entries along each (secondary) diagonal, which is why a Toeplitz matrix is also called a diagonal-constant matrix. In contrast to the compound symmetry matrix, not all off-diagonal elements need to have the same value; this is only required within each secondary diagonal. Then, with
\[\vC_{T}=\bigoplus\limits_{k=0}^{d-1} \vP_{d-k}.\]
the hypothesis of a Toeplitz structure of the covariance matrix can be expressed through $\mathcal H_0^{\vv}(\text{T}):\vC_{T}\vv=\vnull_{p}$.

\section{Extended linear covariance structure model}\label{section:Extended linear covariance structure model} 
Model~\eqref{eq:lcsmodel} is quite general but it allows no relations between the components of $\vtheta$.
This is why some simple structures are not covered, for instance, the one in the subsequent example.
\begin{exmp}\label{beispiel}
Consider the matrix structure given through
$\vV=\diag(\rho,\rho^2,\rho^4,\rho^6...,\rho^{2d})$ with $\rho\in (0,1)$, which is a diagonal matrix with special relations between the components. Because of these multiplicative relations, this structure is not contained in model~\eqref{eq:lcsmodel}.  
\end{exmp}

 Although it is easy to choose the underlying $\vV_0,...,\vV_d$, the connection of the components is tantamount to conditions on $\vtheta$. This is also the case for other structures involving relations between the components of $\vtheta$, where especially the autoregressive structure is popular and important; we will consider this structure in Section~\ref{Autoregressive} below. This makes it nearly impossible to express these structures through  $\mathcal H_0^{\vv}: \vC \vv = \vzeta$.
 Instead, we will test hypotheses from a more general model $\mathcal H_0^{\vv}:\widetilde \vC \tilde\vf(\vv) = \in \R^{m}$
 by using a nonlinear function $\tilde\vf:\R^{p}\to \R^{b}$ together with $\widetilde \vC\in\R^{m\times b}$, $b\in\N$. The following corollary shows that, under some conditions on $\tilde \vf$, this hypothesis can be checked through an adapted ATS.


\begin{Ko}\label{Kof}
Let $\tilde \vf:\R^{p}\to \R^{b}$  be a nonlinear, differentiable function  and $\mathcal H_0^{\vv}: \widetilde\vC \tilde\vf(\vv) = \in \R^{m}$ a null hypothesis. Then with a Jacobian matrix $\vJ_{\tilde\vf}(\vv)\neq \vnull_{p\times b}$ such that $\vJ_{\tilde\vf}(\widehat\vv) \stackrel{\mathcal{P}}\to \vJ_{\tilde\vf}(\vv)$,  this hypothesis can be tested with the help of the test statistic
 \[ATS_{\vv}^{\tilde\vf}(\widehat{\vSigma}_{\dvech})= N\left[  \widetilde\vC\tilde\vf(\widehat \vv) - \right]^\top\left[  \widetilde\vC\tilde\vf(\widehat \vv) - 
\right]/\tr\left(\vJ_{\tilde\vf}(\widehat \vv)\widetilde\vC\widehat{\vSigma}_{\dvech}\widetilde\vC^\top\vJ_{\tilde\vf}(\widehat \vv)^\top\right).\]

Together with corresponding Monte-Carlo quantiles $q_{1-\alpha}^{MC,\tilde\vf}$ or bootstrap quantiles $q_{1-\alpha}^{*,\tilde\vf}$ for  $\alpha\in (0,1)$, this leads to consistent asymptotic level-$\alpha$-tests through $\varphi_{ATS}^{\tilde\vf}:=\ind\{ ATS_{\vv}(\widehat \vSigma_{\dvech}) >q_{1-\alpha}^{MC,\tilde\vf}\}$ and $\varphi_{ATS}^{*,\tilde\vf}=\ind\{ ATS_{\vv}(\widehat \vSigma_{\dvech}) >q_{1-\alpha}^{*,\tilde\vf}\}$.
That is, $\mathbb E(\varphi_{ATS}^{*,\tilde\vf}) \to 1 \cdot \ind_{\mathcal H_1^{\vv}} + \alpha \cdot \ind_{\mathcal H_0^{\vv}}$, where $\mathcal H_1^{\vv} : \tilde \vC \tilde \vf (\vv) \neq \tilde \vzeta$, and identically for $\varphi_{ATS}^{\tilde\vf}$.

\end{Ko}

It remains to constructively find an adequate function $\tilde \vf$ which captures the relation between the components of $\vtheta$.
These may equivalently be translated into
restricting the parameter space $\vTheta$ accordingly.
We will in the following focus on relations between the components which can be written as $ \vf(\vv)=\vf(\dvech(\vV))=\vnull_\ell$, where $\vf:\R^{p}\to \R^{\ell},$ is a temporarily defined auxiliary function, $\ell \in \N$.

For the above-mentioned structure \eqref{beispiel},  a possible function would be given through $\vf:\R^{p}\to \R^{d-1}, \ (x_1,...,x_p)\mapsto (x_1-\sqrt{x_2},x_1-\sqrt[4]{x_3},...,x_{1}-\sqrt[2d]{x_d})$ but also many other choices are possible.

Based on such functions $\vf$, the extended linear covariance structure model is given through
\begin{align}
 \label{eq:elcsmodel}
  \widetilde {\mathcal{V}}=\{\vV(\vtheta)\in \R^{d\times d}:\vV(\vtheta)=\vV_0+\theta_1 \vV_1+\theta_2 \vV_2+...+\theta_q \vV_q,\quad \vtheta \in \widetilde \vTheta\subset  \vTheta\}
 \end{align}
where $\widetilde \vTheta = \{\vtheta \in \vTheta: \vf (\dvech(\vV)) = \vnull_\ell \} \neq \emptyset$ arises from $ \vTheta$ by the restrictions mentioned above. It is obviously an extension of the former Model~\eqref{eq:lcsmodel} which results from the special case $\widetilde \vTheta = \vTheta$. Hence, the new model's name is justified.

Since there are different ways to formulate the same relation between components, the function $ \vf$ is not unique, just as the dimension $\ell$ of the image space.
In any case, it is possible to get the required transformation $\tilde\vf$ using this function $\vf$:

\begin{theorem}\label{TheoremFormulationextended}

   Let $\widetilde \vTheta \subset \vTheta $ be the subset that arises from $ \vTheta$   by the restriction to $ \vf(\vv)=\vf(\dvech(\vV))=\vnull_\ell$, for a function $\vf:\R^{p}\to \R^{\ell}, \ell \in \N$.

 For each covariance matrix structure from model~\eqref{eq:elcsmodel} via  $\widetilde \vTheta$ there exists a  $\widetilde \vC\in\R^{m\times p}$, a $\widetilde{\vf}:\R^p\to \R^m$ and a  $\widetilde\vzeta\in\R^m$ so that the hypothesis of this structure can be formulated by   $\mathcal H_0^{\vv}: \widetilde \vC \widetilde \vf(\vv) = \widetilde \vzeta$ with $m= (p-q)+\ell$.\\
        Moreover, if $\vf$ is continuously differentiable on $\dvech(\mathcal V)$ with a non-vanishing Jacobian matrix, the same holds true for some function $\widetilde{\vf}$ with the above properties.

\end{theorem}
This theorem is based on \Cref{{TheoremFormulation}} together with an adequate construction of the underlying $\tilde\vf$ and $\widetilde \vC$, which is presented in the following. This construction shows not only the existence but also makes clear why $\widetilde{\vf}$ and $\vf$ exhibit similar properties and therefore can also serve as a proof.

\begin{re}\label{re:ftilde}
One constructive way for finding a matrix $\widetilde {\vC}$ and a vector $\widetilde{\vzeta}$ as in the previous theorem is based on the values from \Cref{TheoremFormulation} combined with the function $ \vf$. 
In this way, a possible choice is\[\widetilde\vf(\vv)=\begin{pmatrix}\vv\\
 \vf(\vv)\end{pmatrix},\ \widetilde\vC=\begin{pmatrix}
\vC\\
\vI_{\ell}\end{pmatrix}\text{ and } \tilde \vzeta=\begin{pmatrix}\vzeta\\\vnull_\ell\end{pmatrix},\]
which allows to formulate the hypothesis of having this structure through $\mathcal H_0^{\vv}: \widetilde \vC \widetilde \vf(\vv) = \widetilde \vzeta$.
\end{re}
\setcounter{exmp}{0}

\begin{exmp}[continued]
 On the one hand, the structure of \Cref{beispiel} involves diagonality, which can be checked with a minimal number of rows through $\vC=(\vnull_{p_u\times d},\vI_{p_u})$ and $\vzeta=\vnull_{p_u}$. Furthermore, based on the above-mentioned function $\vf:\R^{p}\to \R^{d-1}, \ (x_1,...,x_p)\mapsto (x_1-\sqrt{x_2},x_1-\sqrt[4]{x_3},...,x_{1}-\sqrt[2d]{x_d})$
we obtain $\widetilde \vf:\R^{p}\to \R^{m}, \ (x_1,...,x_p)\mapsto (x_1,...,x_p,x_1-\sqrt{x_2},x_1-\sqrt[4]{x_3},...,x_{1}-\sqrt[2d]{x_d})$.

Then, with $\widetilde \vzeta=\vnull_m$ and $\widetilde \vC=(\vC^\top,\vI_\ell)^\top$, the hypothesis of having this structure can be formulated through
$\mathcal H_0^{\vv}: \widetilde \vC \widetilde \vf(\vv) = \widetilde \vzeta$. 
\end{exmp}

Again, the previous remark together with \Cref{TheoremFormulation} only leads to one possibility for formulating the null hypothesis, and this need not be the most straightforward or the most sensible choice. 
To exemplify an alternative possibility for finding such specifications, let us now treat the autoregressive structure.

\subsection{Autoregressive}\label{Autoregressive}
The autoregressive structure can be seen to be a special case of a Toeplitz matrix. However, an autoregressive covariance matrix has a proportionality property between the neighbouring (secondary) diagonals, making it a more demanding structure. The matrix depends on just one parameter, $\rho$ with $|\rho|\in(0,1)$, and is given through $(\vV)_{ij}=\rho^{|i-j|}$, which shows that the correlation of the components decreases exponentially with the distance between $i,j \in \{1,\dots,d\}$.
For example, an autoregressive structure can be useful when consecutive components belong to neighbouring measurement points. Moreover, this structure is often used for repeated measurements at different time points since further apart measurements should then have a smaller correlation.  Although sometimes also the case $\rho=0$ is included in the definition of autoregressive covariance matrices, this disagrees with the concept of a (strictly) decreasing correlation.
Consequently, we exclude this simple case of testing for the identity covariance matrix.
As explained by \cite{pEB}, the parameter $\rho$ may, in some settings, also depend on the dimension $d$ of the random vectors. \\

 While the equality along the diagonal and along each secondary diagonal could be tested similarly to the case of a Toeplitz matrix, the above-mentioned proportionality makes a transformation of $\vv$ necessary. 
  To this end, we first define a vector $\va=(a_1,...,a_d)$ with $a_k=1+\sum_{\ell=0}^{k-2} (d-\ell)$, $k=1,...,d$. Then $a_k$ is the component of the vector $\dvech(\vB)$ corresponds to the $(1,k)$th entry $\vB_{1k}$, and therefore to the first component of the $(k-1)$-th secondary diagonal, while the 0-th secondary diagonal is the classical diagonal.


Now, to get a test for the autoregressive structure,  we  define the function %

\[\vg:\R^{p}\to \R^{d-1}, (x_1,...,x_p)\mapsto \left(\frac{\tfrac1{d-1}\cdot\sum\limits_{k=0}^{d-2} x_{a_2+k}}{\tfrac1{d-0}\cdot\sum\limits_{k=0}^{d-1} x_{a_1+k}},
\frac{\tfrac1{d-2}\cdot\sum\limits_{k=0}^{d-3} x_{a_3+k}}{\tfrac1{d-1}\cdot\sum\limits_{k=0}^{d-2} x_{a_2+k}},...,
\frac{\tfrac1{1}\cdot\sum\limits_{k=0}^{0} x_{a_{d}+k}}{\tfrac1{2}\Cdot \sum\limits_{k=0}^{1} x_{a_{d-1}+k}}\right),\]

 which computes ratios of means along neighboring secondary of the covariation matrix.
Under the null hypothesis of an autoregressive structure, each of these ratios would estimate the parameter $\rho$.
Consequently, the proportionality can be checked based on this function.

It is clear that the test is not defined when a secondary diagonal of the empirical covariance matrix has mean zero. But such an empirical covariance matrix would likely indicate that the autoregressive structure does not hold, i.e., the null hypothesis should be rejected.
For testing the null hypothesis of an autoregressive matrix, it remains also to check whether the first $d$ elements of $\vv$ are one and whether all secondary diagonals are constant.
 Using $\vC_{HT}=\bigoplus_{k=1}^{d-1} \vP_{d-k}$, this can be done with a hypothesis matrix $\vC_{AR}=\vI_d\oplus\vC_{HT}\oplus \vP_{d-1}$, which allows the formulation of the null hypothesis of an autoregressive matrix through
\[\mathcal{H}_0(AR): \vC_{AR}\ \begin{pmatrix}
 \vv\\
\vg( \vv)\end{pmatrix} =\begin{pmatrix}\veins_d\\\vnull_{p-1}
\end{pmatrix}.\]
Using $\vJ_{\vg}$, the Jacobian matrix of the function $\vg$,\footnote{The concrete form of $\vJ_{\vg}$ can be found in the supplementary material.}
this structure can be tested based on \Cref{Kof}.\\

For the sake of completeness, we will exemplify in the supplement the use of another function $\vh$, which is based on the roots of the elements on the secondary diagonals.
This underlines the non-uniqueness of the function $\widetilde \vf$. This alternative function will also be used in the simulations in \Cref{Simulations} to demonstrate that different reasonable functions, which meet the conditions of \Cref{TheoremFormulationextended}, could result in a very different practical performance.

\section{Structures of correlation matrices}\label{Structures of correlation matrices}

For positive definite covariance matrices $\vSigma$, the corresponding correlation matrix $\vR$ allows for focusing more on the dependency between the components. Therefore, structures of correlation matrices are also of interest. Since the diagonal elements of a correlation matrix contain no information, another kind of vectorization should be used to prepare hypothesis tests about correlation structures. In \cite{sattler2023correlation}, a so-called
upper-half-vectorization $\vech^-$ is used to formulate hypotheses 
$\mathcal H_0^{\vr}: \widetilde \vC \vech^-(\vR) = \widetilde\vzeta$,
 for the vectorized correlation matrix   $\vech^-(\vR)= (r_{12},\dots,r_{1d},r_{23},\dots,r_{2d},\dots,r_{(d-1)d})^\top\in\R^{p_u}$.
For this hypothesis, tests can be developed  based on the central limit theorem

\[\sqrt{N}(\vech^-(\widehat \vR)-\vech^-(\vR))\stackrel{\mathcal D}{ \longrightarrow}\vZ\sim\mathcal{N}_{p_u}\large(\vnull_{p_u},\vUpsilon\large) \quad \text{as \ } N \to \infty,\]

where the concrete form of the covariance matrix $\vUpsilon\in\R^{p_u\times p_u}$ can be found in Theorem 3.1 of \cite{sattler2023correlation} together with a consistent estimator $\widehat \vUpsilon$.
 Therein, the ATS formulated for these vectorized correlation matrices and approaches to calculating appropriate quantiles can also be found: a classical Monte-Carlo approach, a parametric bootstrap, and a Taylor-based Monte-Carlo approach.
Through the alternative diagonal vectorization, $\vr=\dvech^-(\vR)= (r_{12},r_{23},\dots,r_{(d-1)d},r_{13},\dots,r_{1d})^\top\in\R^{p_u}$,
 all three approaches can again be adapted, using the convergence
 \[\sqrt{N}(\widehat \vr-\vr)\stackrel{\mathcal D}{ \longrightarrow}\vZ\sim\mathcal{N}_{p_u}\large(\vnull_{p_u},\vUpsilon_{\dvech}\large)  \quad \text{as \ } N \to \infty.\]

Here, $ \widehat \vr = \dvech^-(\vR)$ and $\vUpsilon_{\dvech}=\widetilde \vA  \vUpsilon \widetilde \vA^\top$, where $\widetilde \vA \in\R^{p_u\times p_u}$ satisfies $ \dvech^-(\vB)=\widetilde \vA\vech^-(\vB)$ for each matrix $\vB\in \R^{d\times d}$.
One exemplary use of the just-prepared structure is the testing for diagonality of a correlation matrix by $\mathcal{H}_0^{\vr}(D):\vI_{p_u}\vr=\vnull_{p_u}$. 
Similarly to the linear covariance structure model, a linear correlation structure can be defined  with  $q<p_u$ as
\begin{align}
 \label{eq:lcrsmodel}
 \mathcal{R}=\{\vR(\vtheta)\in \R^{d\times d} \! : \! \vR(\vtheta)=\vR_0+\theta_1 \vR_1+...+\theta_q \vR_q, \quad \vtheta = ( \theta_1, \dots, \theta_q)^\top\in\vTheta\subset \R^q\}
\end{align}
where $\vR_{0},...,\vR_q$ are symmetric matrices and matrices and parameter space $\vTheta$ so that $\vR$ is a correlation matrix.
Consequently, $\diag(\vR_0)=\veins_d$ and $\diag(\vR_1)=...=\diag(\vR_q)=\vnull_d$, which is one of the main differences compared to the model ~\eqref{eq:lcsmodel}.

Similarly to~\eqref{eq:elcsmodel}, the extended linear correlation structure can be defined.
\begin{Ko}\label{SameforCorrelation}
The results of \Cref{TheoremFormulation}, \Cref{Kof}  and \Cref{TheoremFormulationextended} hold analogous for correlation matrices and upper triangular vectorization.
\end{Ko}
Now with applications of \Cref{SameforCorrelation}, also the correlation matrix can be tested for the above structures, which is often termed as heterogenous covariance structures.

\subsection{Heterogenous Versions of Structures}
 For so-called heterogeneous variations of structures from \Cref{Linear Covariance Structure Model},
the classical covariance matrix (compound symmetry, Toeplitz, or autoregressive) is multiplied from both sides with a positive definite diagonal matrix, say, $\diag(\iota_1,...,\iota_d)$, where $\iota_1, \dots, \iota_d > 0$.
In this sense, the diagonality of a matrix can also be seen as a heterogenous sphericity structure.\\
It is clear that a heterogeneous Toeplitz structure of the covariance matrix is equivalent to a Topelitz structure of the correlation matrix and, in the same way, for all other mentioned structures. Therefore, the hypotheses regarding heterogeneous versions can be formulated similarly using the vectorized correlation matrix $\vr$.
Consequently, no further conditions on the diagonal elements are necessary, and the hypothesis matrix only requires small modifications.
For the heterogenous Toeplitz structure, the hypotheses can be formulated as  $\mathcal H_0^{\vr}(\text{HT}):\vC_{HT}\vr=\vnull_{p_u}$ with
\[\vC_{HT}=\bigoplus\limits_{k=1}^{d-1} \vP_{d-k},\]
while the hypothesis of a heterogenous compound symmetry structure can be formulated as $\mathcal H_0^{\vr}(\text{HCS}):\vP_{p_u}\vr=\vnull_{p_u}$ and tested with \Cref{SameforCorrelation}.\\

A heterogeneous version also exists for the autoregressive structure, which can be tested with adapted matrices and the vectorized correlation matrix. For testing, the first part of the matrix $\vC_{AR}$ is not required, so that the hypothesis matrix reduces to  $\vC_{HAR}=\vC_{HT}\oplus\vP_{d-2}$.
As a result, we proceed as for the classical autoregressive structure, based on an adapted function

 \[\check\vg:\R^{p_u}\to \R^{d-2}, (x_1,...,x_{p_u})\mapsto \left(
\frac{ \tfrac 1{d-2}\cdot\sum\limits_{k=0}^{d-3} x_{a_3+k-d}}{\tfrac 1 {d-1}\cdot\sum\limits_{k=0}^{d-2} x_{a_2+k-d}},...,
\frac{\tfrac 1 1\cdot\sum\limits_{k=0}^{0} x_{a_{d}+k-d}}{\tfrac 1 2\Cdot \sum\limits_{k=0}^{1} x_{a_{d-1}+k-d}}\right),\]

Then the hypothesis of an heterogeneous autoregressive matrix can be formulated as \[\mathcal{H}_0^{\vr}:\vC_{HAR} \begin{pmatrix}\vr\\ \check\vg(\vr)\end{pmatrix}=\vnull_{p-2}.\]

\section{Combined tests for multiple structures }\label{Combined}
In the previous sections,  hierarchical relationships can be found for most of the structures: for example, each sphericity matrix is also a compound symmetry matrix; each compound symmetry matrix is also a Toeplitz matrix, and so on. Therefore, it would be desirable to allow for subsequent testing about weaker structures whenever a test rejects a hypothesis about a stronger structure. This is possible with a multiple contrast test; see, e.g., Section~6 of \cite{sattler2023correlation} and \cite{munko2023b} for an appropriate bootstrap technique. 
For didactic reasons, we will outline the approach of a combined test with the sphericity structure as an example.

Sphericity can be understood as diagonality with equal diagonal elements, which is also apparent from the composition of the hypothesis matrix $\vC_{S}=\vP_d\oplus \vI_{p_u}$ in comparison to the hypothesis matrix for testing diagonality, $\vC_D=\vnull_{d\times d}\oplus \vI_{p_u}$.
Now, for preparing the multiple contrast test, we define
\[\vT=\begin{pmatrix}
T_1\\
\vdots\\
T_p
\end{pmatrix}= \sqrt{N}\begin{pmatrix}
\vc_{S\bullet 1} \widehat \vv\\\vdots\\
\vc_{S\bullet p} \widehat \vv
\end{pmatrix}=\sqrt{N}\vC_S \widehat \vv,\]
where $\vc_{S\bullet k}, k=1,\dots, p,$ is the $k$-th row vector of $\vC_S$.
Under the null hypothesis of a sphericity matrix, this statistic fulfils $\vT\stackrel{\mathcal{D}}{\to}\mathcal{N}_p(\vnull_p,\vC_S\vSigma_{\dvech}\vC_S^\top)$

as $N \to \infty$. One possibility for mimicking this convergence is by means of a computer-intensive parametric bootstrap scheme. That is, based on the parametric bootstrap sample $\vY_1^*$,...,$\vY_N^*\stackrel{i.i.d.}{\sim}\mathcal{N}_p(\vnull,\widehat \vSigma_{\dvech})$ which is normally distributed conditionally on the data, a bootstrap version of the test statistic can be defined through
\[\vT^{1,*}=\begin{pmatrix}
T_1^{1,*}\\
\vdots\\
T_p^{1,*}
\end{pmatrix}= \sqrt{N}\begin{pmatrix}
\vc_{S\bullet 1} \overline \vY^*\\\vdots\\
\vc_{S\bullet p} \overline \vY^*
\end{pmatrix}
=\sqrt{N}\vC_S\overline \vY^* \]
with $\overline \vY^*=N^{-1}\cdot \sum_{k=1}^N  \vY_k^*$. Since, this approach in \cite{sattler2022} was developed for multi-group settings, it could be simplified here to save computation time.

It is easy to see that $\vT^{1,*}$ has the same asymptotic distribution as $\vT$ under the null hypothesis. Repeating this $B$ times, i.e., there are $B$ conditionally independent realizations of bootstrap samples, this leads to $\vT^{1,*},...,\vT^{B,*}$.
Denote by $q_{\ell,\beta}^*$ the empirical $(1-\beta$) quantile  for $|T_\ell|$, $\ell=1,...,p$, conditionally on the original data. To find an appropriate level $\beta$ to control the family-wise type-I error rate, we calculate similarly as in \cite{munko2023a} 

\[\widetilde \beta= \max\left(\beta \in \left\lbrace 0,\frac{1}{B},..., \frac{B-1}{B}\right\rbrace\Big\lvert \frac{1}{B} \sum\limits_{b=1}^B \max\limits_{\ell=1,...,p}\left(\ind\left(|T_\ell^{b,*}|>q_{\ell,\beta}^{*}\right)\right)\leq \alpha\right),\]
i.e., the maximum local level, which results in a global level of $\alpha$.

Then, the null hypothesis of sphericity can be rejected if and only if \[\max\limits_{\ell=1,...,p}\left(\ind\left( |T_\ell|>q_{\ell,\widetilde \beta}^{*}\right)\right)=1 \quad \Leftrightarrow \quad \max\limits_{\ell=1,...,p} \left(\frac{|T_\ell|}{q_{\ell,\widetilde \beta}^{*}}\right)>1\]
where we define $0/0:=1$. It is easy to see that this procedure leads to a test for this covariance structure, which has the asymptotic level $\alpha$, as $N\to \infty$. 
Since, for each component of the vector-valued test statistic $\vT$, the same local level $\widetilde \beta$ is used, each component has the same weight and importance.

Now, if $\max( |T_{1 }|/q_{1,\widetilde \beta}^*,....,|T_{d }|/q_{d,\widetilde \beta}^*)> 1$,  then the hypothesis of diagonality is rejected too. In cases with $\max( |T_{d+1 }|/q_{d+1,\widetilde \beta}^*,....,|T_{p }|/q_{p,\widetilde \beta}^*)> 1$ but $\max( |T_{1 }|/q_{1,\widetilde \beta}^*,...., |T_{d }|/q_{d,\widetilde \beta}^*)\leq 1$, the hypothesis of sphericity is rejected, but the hypothesis of diagonality is not.

With another hypothesis matrix, based on two appropriate Tukey-type contrast matrices (see, e.g. \cite{tukey1953}), also further conclusions would be possible to see which components of the vectorized covariance matrix have caused the rejection.

Often, tests based on the maximum of multiple contrasts are based on so-called equicoordinate quantiles (see e.g. \cite{rubarth2022}) instead of bootstrap quantiles. But this requires the diagonal elements of $\vC_S\vSigma_{\dvech}\vC_S^\top$ to be positive. 
However, this requirement need not be met in general, at least not without further restrictions on $\vSigma_{\dvech}$.

\section{Simulation study}\label{Simulations}
In this section, we will investigate the performance of some of the tests proposed in the previous sections.
In particular, in \Cref{TypI}, we will simulate the actual Type-I error probability of tests with a nominal level of $\alpha = 5\%$; in \Cref{TypII}, we will assess the power of some of the tests. \\
For practical use, we developed functions in the R-computing environment (\cite{R}),
for all our test procedures. The corresponding source code, together with instructions to use the package, can be found in our GitHub repository.\footnote{\url{https://github.com/PSattlerStat/StructureTestCovCorrMatrix}}
\subsection{Type-I error}
\label{TypI}
Since diagonality was already tested in \cite{sattler2023correlation}, we now focus on testing for an autoregressive and a Toeplitz structure based on the above-introduced matrices. 
 Hence, we have one structure based on just one parameter and one based on $d$ parameters, allowing for interesting comparisons. \cite{herzog2007} assumed a relation between the number of parameters and the required sample size for a sufficient approximation of the asymptotic distribution; we would like to investigate whether this is also the case for our novel tests.
 Based on results from \cite{sattler2023hypothesis}, we use the hypothesis matrix formed by removing zero-rows from the existing unique projection matrix for both kinds of covariance matrix structures.
For the autoregressive structure, we chose the parameter $\rho=0.65$ and get $\vV_1$; for the Toeplitz matrix, we chose the covariance matrix \[\vV_{2}=\begin{pmatrix}1.2&0.9&0.8&0.4&0.1\\
         0.9 &1.2 &0.9 &0.8 &0.4\\
         0.8 &0.9 &1.2 &0.9 &0.8\\
         0.4 &0.8 &0.9 &1.2 &0.9\\
         0.1 &0.4 &0.8 &0.9 &1.2\end{pmatrix}.\] For both covariance matrices $\vV_1$ and $\vV_2$, we simulate five dimensional random vectors $\vep_{j}=\vV^{1/2}\vZ_{j}$ with the entries $Z_{j1},...,Z_{j5}$ of $\vZ_j$ being i.i.d.\ and following

\begin{itemize}
\item a standard normal distribution, 
\item a standardized centered gamma distribution,
\item a standardized centered skew normal distribution with location parameter $\xi=0$, scale parameter $\omega=1$ and $\gamma=4$. The density of a skew normal distribution is given through $x \mapsto \frac{2}{\omega} \varphi\left(\frac{x-\xi}{\omega}\right)\Phi\left(\gamma\left(\frac{x-\xi}{\omega}\right)\right)$, where $\varphi$ denotes the density of the standard normal distribution and $\Phi$ the according distribution function,
\item or a standardized centred $t$-distribution with 9 degrees of freedom,
\end{itemize}
while the sample sizes are $\vN \in \{25,50,100,250\}$. Based on the results from \cite{sattler2022}, we consider the ATS with parametric bootstrap and with Monte-Carlo-based critical values. 

To this end, $\widehat{\vSigma}_{\dvech}$ is used to estimate the eigenvalues $\lambda_1,...,\lambda_m$ of $(\vC\vSigma\vC^\top)$ through $\hat \lambda_1,...,\hat \lambda_m$. This allows us to generate realizations of $\sum_{\ell=1}^m \hat\lambda B_\ell$ with $B_1,...,B_m\stackrel{i.i.d.}{\sim}\chi_1^2$, which asymptotically coincides with the distribution of $ATS_{\vv}$ under the null hypothesis.
We use 1,000 bootstrap runs and 10,000 Monte-Carlo steps. The empirical type-I error rates, based on 10,000 simulation runs, can be seen in Tables~\ref{tab:toeplitzstruktur} and~\ref{tab:autostruktur}. Both approaches, i.e., those based on the functions $\vh$ and $\vg$, are used for testing the autoregressive structure. 
In the just-mentioned tables, empirical type-I error rates inside the $95\%$ binomial interval $[0.0458; 0.0543]$ are printed in bold-type.

For testing whether the covariance matrix is a Toeplitz matrix, the parametric bootstrap has a better small sample performance than the Monte-Carlo approach. For larger sample sizes, the type-I error rates of the two approaches get closer to each other.  
Excepting the gamma distribution scenario, all simulated type-I error rates of the parametric bootstrap are contained in the 95\% binomial interval around $\alpha = 5\%$, while both tests fulfil Bradley's liberal criterion \citep{bradley1978} in all cases, and they exhibit small type-I error rates in general, especially for the $t_9$ distribution.
Applied statisticians often consulate this criterion, for example in quantitative psychology.  It states a procedure as 'acceptable' if it has a type I error rate between $0.5\alpha$ and $1.5\alpha$.

  \begin{table}[htbp]
   
\begin{center}
    \begin{tabular}{x{1.5pt}lx{1.5pt}c|c|c|cx{1.5pt}c|c|c|cx{1.5pt}}\specialrule{1.5pt}{0pt}{0pt}
 \rowcolor{ashgrey}  
      &\multicolumn{4}{|cx{1.5pt}}{ATS-Para}&\multicolumn{4}{cx{1.5pt}}{ATS}\\\specialrule{1.5pt}{0pt}{0pt} 
\rowcolor{gainsboro}N&25&50&100&250&25&50&100&250
       \\\specialrule{1.5pt}{0pt}{0pt}
\rowcolor{lightyellow}$t_9$ & \bf{4.72} & \bf{4.75} & \bf{5.00} & \bf{4.70} & \bf{5.36} & \bf{4.99} & \bf{5.10} & \bf{4.66} \\ 
 \rowcolor{gainsboro}Normal  & \bf{5.20} & \bf{5.26} & \bf{5.15} & \bf{4.82} & 5.88 & 5.62 & \bf{5.24} & \bf{4.81} \\ 
  \rowcolor{lightyellow}  Skew normal  & \bf{5.15} & \bf{5.28} & \bf{4.84} & \bf{5.37 }& 5.91 & 5.49 & \bf{4.95} & \bf{5.39} \\ 
 \rowcolor{gainsboro}Gamma & 4.27 & 4.29 & 4.36 & 4.51 & \bf{4.93} & \bf{4.60} & 4.36 & 4.55 \\  \specialrule{1.5pt}{0pt}{0pt}

\end{tabular}
  \caption{Simulated type-I-error rates ($\alpha=5\%$) in percentage points, for testing  Toeplitz covariance matrix structures, with ATS based on parametric bootstrap (ATS-Para) and based on Monte-Carlo approach (ATS). The observation vectors have dimension 5, covariance matrix $\vV_{2}$, and different distributions and sample sizes are considered. Rejection rates within the $95\%$ confidence interval around $\alpha=5\%$ are printed in bold-type.}
    \label{tab:toeplitzstruktur}
    \end{center}
 
    \end{table}

For the considerably more challenging hypothesis of an autoregressive structure the results are less good, which is not  surprising because the proportionality, which is the only difference to the Toeplitz matrix, is difficult to check across all secondary diagonals.
But for the bootstrap approach applied based on the function $\vg$, the liberal criterion is fulfilled for all considered distributions, while for $N>50$ almost all error rates also are in the binomial interval. The Monte-Carlo approach however is, in this case too conservative and also needs sample sizes of $N>50$ to fulfill Bradleys liberal criterion.\\
In comparison, for the function $\vh$, the parametric bootstrap is too liberal and has slightly worse results than for the function $\vg$, especially for larger sample sizes. In contrast, the Monte-Carlo approach based on $\vh$ is too conservative but substantially better than for $\vg$, which makes it applicable for large sample sizes.\\

 Although these results  make the $\vg$ function a more reasonable choice, other functions and hypothesis matrices may exist for the autoregressive structure with better performance than those based on the function $\vg$.\\

\begin{table}[htb]
  \begin{center}
    \begin{tabular}{x{1.5pt}lx{1.5pt}c|c|c|cx{1.5pt}c|c|c|cx{1.5pt}}\specialrule{1.5pt}{0pt}{0pt}
   \rowcolor{ashgrey}      &\multicolumn{4}{|cx{1.5pt}}{ATS-Para-$\vg$}&\multicolumn{4}{cx{1.5pt}}{ATS-$\vg$}\\ \specialrule{1.5pt}{0pt}{0pt} 
\rowcolor{gainsboro}N&25&50&100&250&25&50&100&250
       \\\specialrule{1.5pt}{0pt}{0pt}
\rowcolor{lightyellow}$t_9$ & 3.98 & 3.86 & \bf{4.62} & \bf{4.95} & 1.76 & 1.89 & 2.61 & 3.67 \\   \hline
\rowcolor{gainsboro}Normal   & 3.80 & 3.94 & \bf{4.70} & \bf{4.97} & 1.67 & 1.87 & 2.49 & 3.81 \\   \hline
\rowcolor{lightyellow}  Skew normal & 4.37 & 4.21 & 4.37 & \bf{5.00} & 1.79 & 2.03 & 2.51 & 3.62 \\   \hline
\rowcolor{gainsboro}Gamma  & \bf{5.15} & \bf{5.00} & \bf{4.66} & \bf{4.70} & 2.70 & 2.62 & 2.75 & 3.64 \\    \specialrule{1.5pt}{0pt}{0pt}
\multicolumn{9}{x{1.5pt}cx{1.5pt}}{}\\\specialrule{1.5pt}{0pt}{0pt}
   \rowcolor{ashgrey}     &\multicolumn{4}{|cx{1.5pt}}{ATS-Para-$\vh$}&\multicolumn{4}{cx{1.5pt}}{ATS-$\vh$}\\\specialrule{1.5pt}{0pt}{0pt}  
\rowcolor{gainsboro}N&25&50&100&250&25&50&100&250
       \\\specialrule{1.5pt}{0pt}{0pt}
\rowcolor{lightyellow}$t_9$ &5.50 & \bf{5.33} & \bf{4.96} & \bf{4.93} & 3.50 & 3.29 & 3.99 & \bf{4.63} \\   \hline
\rowcolor{gainsboro}Normal   &5.61 & 5.82 & 5.54 & 5.58 & 3.27 & 3.57 & 4.18 & \bf{4.87} \\   \hline
 \rowcolor{lightyellow} Skew normal&  6.49 & 5.92 & 5.47 & \bf{5.41} & 3.93 & 3.91 & 4.13 & \bf{4.90} \\   \hline
\rowcolor{gainsboro}Gamma & 6.75 & 5.69 & \bf{4.83} & \bf{4.65} & \bf{4.68} & 3.87 & 3.79 & 3.99 \\    \specialrule{1.5pt}{0pt}{0pt}

\end{tabular}
\end{center}
  \caption{Simulated type-I-error rates ($\alpha=5\%$) in percentage points, for testing autoregressive covariance matrix structures, with ATS based on parametric bootstrap (ATS-Para-$\vg$ and ATS-Para-${\vh}$) and based on Monte-Carlo simulation (ATS-$\vg$ and ATS-$\vh$). The observation vectors have dimension 5, covariance matrix $(\vV_1)_{ij}=0.65^{|i-j|}$ and different distributions and sample sizes are considered. Rejection rates within the $95\%$ confidence interval around $\alpha=5\%$ are printed in bold-type.}
    \label{tab:autostruktur}

    \end{table}

This section shows that with our approach, hypotheses regarding the structure or pattern of the covariance matrix can be tested with a suitable hypothesis matrix with a satisfactory type-I error control. Some adaptations must be made for some more complex structures, and, in general, a large sample size is recommended for reliable results.

\subsection{Power}
\label{TypII}
Besides the type-I-error rate, the capability to recognize deviations from the null hypothesis is a crucial criterion for a good test. Here, we will focus on testing the null hypothesis of an autoregressive covariance matrix, which is the most demanding structure. Since an autoregressive structure is a special case of a Toeplitz matrix, we will check whether the test rejects the null hypothesis when the considered matrix is a Toeplitz matrix without the autoregressive structure. To this end, we let  $\vV_\delta=(1-\delta) \vV_1+ \delta\vV_2, \delta \in[0,1]$, be the true covariance matrices,  where we re-used the
 autoregressive matrix $\vV_1$ and the Toeplitz matrix $\vV_2$  from the previous section.
 That is, we consider mixtures of an autoregressive and a classical Toeplitz matrix under the alternative hypothesis. Now, for $\delta=0,0.1,...,1$, we generate $N=250$ five-dimensional normally distributed random vectors with covariance matrix $\vV_\delta$ and repeat this 1,000 times to estimate the power.
 
The results are given in \Cref{tab:PAR1}, and we see that the power increases fast when leaving the autoregressive structure.
Since the autoregressive structure is a subset of the Toeplitz structure, we consider the displayed power to be very good, 
particularly because the diagonal elements of $\vV_2$ are close to 1. We also noticed a slight power advantage of the bootstrap-based test over the Monte-Carlo-based test, which is not surprising because of the earlier mentioned conservativeness.

   \begin{table}[htbp]
   \begin{small}

\begin{center}
    \begin{tabular}{x{1.5pt}lx{1.5pt}r|r|r|r|r|r|r|r|r|r|rx{1.5pt}}\specialrule{1.5pt}{0pt}{0pt}

\rowcolor{ashgrey}  \hspace{0.7cm}$\delta$&0.0&0.1&0.2&0.3&0.4&0.5&0.6&0.7&0.8&0.9&1.0
       \\\specialrule{1.5pt}{0pt}{0pt}
\rowcolor{lightyellow}ATS-Para-$\vg$ & 4.8 & 6.9 & 12.4 & 24.2 & 40.1 & 64.0 & 80.6 & 91.8 & 97.5 & 99.6 & 100.0 \\  \hline
\rowcolor{gainsboro}ATS-$\vg$ & 3.5 & 5.1 & 10.0 & 19.6 & 35.0 & 57.3 & 76.2 & 89.5 & 96.8 & 99.6 & 99.8 \\  \specialrule{1.5pt}{0pt}{0pt}

\end{tabular}
  \caption{Simulated power ($\alpha=5\%$) in percentage for testing autoregressive covariance matrix structures, with ATS based on parametric bootstrap and Monte-Carlo simulation and $N=250$}.
    \label{tab:PAR1}
    \end{center}
     \end{small}
     \end{table}

The power simulation results for another underlying distribution and also for the approach based on the function $\vh$ is part of the supplementary material. 
There, it can be seen that the use of $\vh$ results in a substantially worse type-I error control which also makes $\vg$ appear preferable.

\section{\textsc{Illustrative Data Analysis}}\label{Illustrative Data Analysis}

After the assessment of the reliability of the tests in small sample regimes in the previous section, we will now illustrate their application to the EEG-data set from the \textsc{R}-package \textit{manova.rm} (\cite{manova}).
$N=160$ patients with three different diagnoses of impairments (subjective cognitive complaints  (SCC), mild cognitive impairment (MCI), and Alzheimer's disease (AD)) participated in a trial conducted at the University Clinic of Salzburg, Department of Neurology.
Thereby, neurological parameters such as the z-score of the brain rate and the Hjorth complexity were measured at three different locations of the head: temporal, frontal, and central. \Cref{tab:EEG1} contains the numbers of patients according to sex and diagnosis.

\begin{table}[h]
\centering
\begin{tabular}{l|c|c|c|}
&AD&MCI&SCC\\
\hline
male&12&27&20\\
\hline
female&24&30&47\\
\hline
\end{tabular}
\caption{numbers of observations for different factor level combinations of sex and diagnosis.}\label{tab:EEG1}
\end{table}

In this analysis, we want to investigate whether the position of the measuring points influences the measured values. Similar questions are often considered in repeated measures designs, where the repetitions have a temporal context, to investigate whether there is a time effect. One way to check such an impact is to compare the means of the three locations and use, for example, a one-sample Hotelling's $T^2$ test \citep{anderson2003}. However, this is not the only way the position of the measurement points can influence the measurements. It could also affect the variance of the individual measure points, as well as the dependency structure between them.
 
 \begin{table}[htp]
\centering
\begin{tabular}{x{1.5pt}llx{1.5pt}cx{1.5pt}cx{1.5pt}cx{1.5pt}}
\specialrule{1.5pt}{0pt}{0pt}

\rowcolor{ashgrey} 
Brain rate\hspace*{-0.2cm}&\hspace*{-0.4cm}&Hotelling's $T^2$& ATS-Para for $\mathcal{H}_0^{\vv}$(\text{CS})& ATS-Para for $\mathcal{H}_0^{\vv}$(\text{T})\\

\rowcolor{ashgrey} &\hspace*{-0.215cm}&p-value &p-value  &  p-value  \\
\specialrule{1.5pt}{0pt}{0pt}
\rowcolor{gainsboro}male&\hspace*{-0.215cm}AD &0.9881 & 0.4056 & 0.4883 \\  \hline
\rowcolor{lightyellow} male&\hspace*{-0.215cm}MCI& 0.7472 & 0.4869 & 0.5882 \\ \hline
\rowcolor{gainsboro}male&\hspace*{-0.215cm}SCC & \bf{0.0162} & 0.2380 & 0.2395 \\ \hline
\rowcolor{lightyellow} female&\hspace*{-0.215cm}AD &  0.6483 & 0.5845 & 0.5553\\ \hline
\rowcolor{gainsboro}female&\hspace*{-0.215cm}MCI& 0.9261 & 0.8014 & 0.7572  \\ \hline
\rowcolor{lightyellow}female&\hspace*{-0.215cm}SCC & 0.9391 & 0.6938 & 0.6634\\ \hline

\specialrule{1.5pt}{0pt}{0pt}\end{tabular}
\caption{p-values of one sample Hotelling's $T^2$ test and $\varphi_{ATS}^*$ to check whether the covariance matrix has a compound symmetry and a Toeplitz structure, respectively, applied to the brain rate data.}\label{EEGResultateStruktur2}

\end{table}

Therefore, we also want to consider the covariance matrix and test whether the covariance matrix has a compound symmetry structure. A rejection of this structure allows the conclusion that the variances are different or the correlations between the locations are different. It should be pointed out that the locations are not exchangeable, which is seen from an influence of the measuring point's position. For completeness, we are also testing whether the covariance matrix has a Toeplitz structure. In contrast to a compound symmetry matrix, this would mean that there are systematic differences in the correlations. Such a structure might make sense to represent distances between measurement points; in the present case, however, a Toeplitz structure seems unlikely because all locations are neighbouring.
Due to the low dimension of the measurements ($d=3$ and therefore $p=6$ for the covariance matrix), we expect reliable results from our test, even for the sample size of only  12 observations.

For completeness, we have also applied the one sample Hotelling's $T^2$ test for analyzing differences in the means.  Thereto, we multiply the data with 
\[\vC=\begin{pmatrix}
    1&-1&0\\
    0&  1 & -1
\end{pmatrix}\]
 to check whether $\vC\vmu=\vnull_3 \Leftrightarrow \mu_1=\mu_2=\mu_3$, and therefore investigate a potential influence of a location parameter.

This test is based on a $\chi^2$ distribution and it is available, for example, through the R-package \textit{ICSNP} (\cite{ICSNP}). 

For testing both covariance structures, the compound symmetry and Toeplitz, we used the ATS with the parametric bootstrap based on 10,000 bootstrap runs and calculated the p-values. The results, together with the results of the one-sample Hotelling's $T^2$ test, are displayed in Tables~\ref{EEGResultateStruktur2} and~\ref{EEGResultateStruktur3}.

\begin{table}[htp]
\centering

\begin{tabular}{x{1.5pt}llx{1.5pt}cx{1.5pt}cx{1.5pt}cx{1.5pt}}
\specialrule{1.5pt}{0pt}{0pt}

\rowcolor{ashgrey} Hjorth \hspace*{-.4cm}&\hspace*{-0.45cm}&Hotelling's $T^2$& ATS-Para for $\mathcal{H}_0^{\vv}$(\text{CS})& ATS-Para for $\mathcal{H}_0^{\vv}$(\text{T})\\


\rowcolor{ashgrey}complexity\hspace*{-0.22cm}&\hspace*{-0.45cm}&p-value &p-value  &  p-value  \\
\specialrule{1.5pt}{0pt}{0pt}
\rowcolor{gainsboro} male&\hspace*{-0.45cm}AD &0.4372 & 0.4029 & 0.4110  \\ \hline
\rowcolor{lightyellow}male&\hspace*{-0.45cm}MCI& 0.1276 & 0.1113 & 0.1142 \\ \hline
\rowcolor{gainsboro}male&\hspace*{-0.45cm}SCC & 0.1273 & 0.1412 & 0.1453\\ \hline
\rowcolor{lightyellow}female&\hspace*{-0.45cm}AD & 0.3139 & 0.3809 & 0.3491  \\ \hline
\rowcolor{gainsboro} female&\hspace*{-0.45cm}MCI& 0.9328 & 0.1172 & 0.1207 \\ \hline
\rowcolor{lightyellow}female&\hspace*{-0.45cm}SCC &\bf{ 0.0213} & \bf{0.0073} & \bf{0.0079}\\

\specialrule{1.5pt}{0pt}{0pt}\end{tabular}
\caption{p-values of one sample Hotelling's $T^2$ test and $\varphi_{ATS}^*$ to check whether the covariance matrix has a compound symmetry and a Toeplitz structure, respectively, applied to the Hjorth complexity data.}\label{EEGResultateStruktur3}

\end{table}

At level $\alpha=5\%$, for the brain rate in the group of men with SCC, a difference in the means could be verified, while for the covariance, no structures were rejected. 
In contrast, for the Hjorth complexity, the location's influence could be established for the mean and the covariance for women with SCC since both structures were rejected at level $\alpha=5\%$. 

{To investigate this group in more detail,  further hypotheses could be tested, as equal variances of all components, which were also considered in \cite{sattler2022}. Rejection of this or similar larger null hypotheses would allow us to better understand the location's influence on the covariance matrix.}
Since the compound symmetry structure is a special case of the Toeplitz structure, it is not surprising that the tests for compound symmetry resulted in lower p-values in most groups.

All in all, the noticeable differences in p-values between the mean and covariance matrix structure showed that, for verifying an effect, both aspects should be taken into account: possible differences in the means and the covariance matrices. This example illustrates how both kinds of hypotheses can be used to investigate two aspects of the same question.

\section{Conclusion}
\label{conclude}
In the present paper, we have introduced an approach for testing a multitude of covariance matrix structures or correlation matrix structures.
The same general test procedure can be used for many of the most important members of the linear covariance structure model, by changing only the hypothesis matrix. On the other hand, some basic methods must be used when more complex covariance structures are to be tested which exhibit special dependencies between the components.

This allowed us to use our tests for many hypotheses that were hard to verify otherwise.
At the same time, our approach also covers covariance structures for which tests already exist; hence, it impresses through its applicability to a wide range of structures and patterns.
Also, a combined test for nested structures was introduced, allowing the testing of multiple linked structures simultaneously.

 Although the testing of covariance matrix structures is commonly acknowledged to be a challenging task, our simulations showed that our tests based on bootstrap and Monte-Carlo techniques provide good statistical results.

Some less common covariance patterns are treated in other works, e.g., \cite{steiger1980}, where a so-called circumplex hypothesis is checked, which tests for a Toeplitz matrix with diagonal elements one or a so-called equicorrelation hypothesis, which means that all non-diagonal elements are equal. The example shows that most of the not-considered patterns are related to the treated structures and, therefore, can be investigated similarly. 

Although every covariance (or correlation) matrix from the linear covariance structure model can be tested based on this approach, 
there exist structures which are not appropriately described by this model.  One of these is the {factor analysis model}
given through $\vV=\vL\vL^\top+\vD$, with $\vL\in \R^{d\times k}$ and a diagonal matrix $\vD\in\R^{d\times d}$ (see, e.g. \cite{lawley1973}).
Even for $k=1$, a possible transformation and hypothesis matrix would be too complex while, for greater $k$, the relation between the components of $\vV$ cannot be expressed through our approach based on $\vv$. As a consequence, we will try to generalize our results to other classes of structures in future research; for example, the linear inverse covariance structure model from \cite{jensen1988}.

In future work, we will develop an R-package to promote the usage of the introduced tests.
Also, we plan to expand the class of testable covariance structures further. Combined hypotheses of structure and equality of covariance matrices, as considered in \cite{mckeown1996}, are another aspect for further research.

\section*{Acknowledgment}
Paavo Sattler would like to thank the German Research Foundation for the support received within project  PA 2409/4-1. Moreover, the authors also want to thank Manuel Rosenbaum for helpful discussions.

\section*{Supplementary material available online}
The supplementary material contains all proofs and additional simulation results.

\appendix

\part*{Appendix}\label{Appendix}
In the following, some more cumbersome matrices have to be defined, where we use the previously defined vector $\va=(a_1,...,a_d)$ containing the indices of components in the half-vectorized matrix $\dvech(\vB)$, which belong to the first element of the corresponding secondary diagonal. So for  $k=1,...,d$ the number $a_k=1+\sum_{\ell=0}^{k-2} (d-\ell)$ gives the index of the first element of the (k-1)-th secondary diagonal.
 Together with $\ve_{k,d} \in \mathbb{R}^d$ denoting the $k$-th standard basis vector\footnote{the $d$-dimensional vector with value one in the $k$-th component and zeros elsewhere}, all required  functions and matrices can be expressed , as for example the transformation matrix~$\vA$.

\section{Transformation from \textbf{$\vech$} to \bf{$\dvech$}}\label{Transform}
Since $\dvech$ is only a rearrangement of the elements of $\vech$ in another order, there is a one-to-one relation between both vectorizations, given through $\vA\vech(\vB)=\dvech(\vB)$ with 
\[\vA=\sum\limits_{\ell=0}^{d-1}\sum\limits_{k=1}^{d-\ell}\ve_{a_{\ell+1}+k-1,p}\cdot \ve_{{a}_{k}+\ell,p}^\top.\]
Therefore from \Cref{Nv1} it directly follows that $$\sqrt{N} (\widehat \vv -\vv)=\sqrt{N} \vA(\vech(\widehat\vV) -\vech(\vV))\stackrel{\mathcal {D}}{\longrightarrow} {\mathcal{N}_{p}\left(\vnull_{p},\vSigma_{\dvech} \right)}$$
as $N\to\infty$, with $\vSigma_{\dvech}=\vA\vSigma\vA^\top=\Cov(\dvech(\vepsilon_{1}\vepsilon_{1}^\top))$.

 With a consistent estimator $\widehat \vSigma$ for the unknown covariance matrix $\vSigma$ and the continuous mapping theorem, $\vSigma_{\dvech}$ can be estimated through  $\widehat \vSigma_{\dvech}=\vA\widehat\vSigma\vA^\top$. Therefore, all methods for calculating quantiles from \cite{sattler2022} can be used here, but we will sketch them here for didactical reasons.
For the Monte-Carlo approach, we use that, under the null hypothesis, 
$$ATS_{\vv}\stackrel{\mathcal{D}}{\to}\sum_{k=1}^p \lambda_k B_k$$
as $N\to\infty$, with $B_k\stackrel{i.i.d.}{\sim}\chi_1^2$ and $\lambda_1,...,\lambda_p$ are the eigenvalues of $(\vC \vSigma_{\dvech}\vC^\top)/\tr(\vC \vSigma_{\dvech}\vC^\top)$.
These eigenvalues can be estimated with the help of $\widehat \vSigma_{\dvech}$, and, with realizations of $B_1,...,B_p\stackrel{i.i.d.}{\sim}\chi_1^2$, one (approximated) realization of the previously displayed weighted sum can be computed. Repeating this procedure independently, for example, 10,000 times, an empirical quantile $q_{1-\alpha}^{MC}$ of the estimated weighted sums can be calculated and used for the test.\\

All of these arguments similarly hold for $\dvech^-$ and  $\vech^-$ with
\[\widetilde \vA=\sum\limits_{\ell=0}^{d-1}\sum\limits_{k=1}^{d-\ell}\ve_{a_{\ell+1}+k-1,p_u}\cdot \ve_{{\va}_{k}+\ell,p_u}^\top,\]but we reiterate this for a better understanding.
In \cite{sattler2023correlation}, the following connection between the vectorized empirical covariance matrix and the vectorized empirical correlation matrix was shown:
\begin{equation}\sqrt{N}(\vech^-(\widehat \vR)-\vech^-(\vR))=\vM(\vech(\vV),\vech^-(\vR))\Cdot \sqrt{N} (\vech(\widehat \vV)-\vech(\vV))+\lan_P(1). \end{equation}
Using this matrix-valued function  $\vM(\vech(\vV),\vech^-(\vR))$ ,
it follows that \[\sqrt{N}(\vech^-(\widehat \vR)-\vech^-(\vR))\stackrel{\mathcal D}{ \longrightarrow}\mathcal{N}_{p_u}\large(\vnull_{p_u},\vUpsilon\large)\]
with $\vUpsilon=\vM(\vech(\vV),\vech^-(\vR))\vSigma \vM(\vech(\vV),\vech^-(\vR))^\top$.
Again with the usage of $\widetilde \vA\vech^-(\vB)=\dvech^-(\vB)$, $\dvech(\vR)=\vr$, and $\dvech(\widehat\vR)=\widehat\vr$, we get
 \[\sqrt{N}(\widehat \vr-\vr)\stackrel{\mathcal D}{ \longrightarrow}\mathcal{N}_{p_u}\large(\vnull_{p_u},\vUpsilon_{\dvech}\large)\]
with $\vUpsilon_{\dvech}=\widetilde \vA\vUpsilon\widetilde \vA^\top$.

\section{Bootstraps}\label{Bootstraps}
For a parametric bootstrap technique, two different approaches are possible. For the first one, similarly as in \Cref{Transform}, a bootstrap sample is generated with conditionally independent $\vY_1^*,...,\vY_{N}^*\sim\mathcal{N}_p(\widehat \vv,\widehat\vSigma_{\dvech})$. The difference is the expectation vector, which is necessary for the following idea:
It is clear from \cite{sattler2022} that, given the data,
\[\sqrt{N}(\overline \vY^*- \widehat \vv)\stackrel{\mathcal D}{\longrightarrow}\mathcal{N}_{p}(\vnull_p,\vSigma_{\dvech} )\]
holds in probability
and therefore, owing to the multivariate delta method, 
\[\sqrt{N}\vC (\vf(\overline \vY^*)- \vf(\widehat \vv))\stackrel{\mathcal D}{\longrightarrow}\mathcal{N}_{m}(\vnull_m,\vC\vJ_{\vf}(\vv)\vSigma_{\dvech}  \vJ_{\vf}(\vv)^\top\vC^\top)\]
holds in probability as well.
Now, a consistent variance estimator for the bootstrap test statistic remains to be found. Since $$\E(\vY_1^*|\vX_1, \dots, \vX_N)=\widehat \vv\stackrel{\mathcal{P}}{\to}\vv \quad \text{and} \quad  \Cov(\vY_1^*|\vX_1, \dots, \vX_N)=\widehat\vSigma_{\dvech}\stackrel{\mathcal{P}}{\to}\vSigma_{\dvech},$$ we can use $\vJ_{\vf}(\overline \vY^*)\widehat\vSigma_{\dvech}^*  \vJ_{\vf}(\overline \vY^*)^\top$ as a variance estimator, while $\widehat\vSigma_{\dvech}^*$ denotes the empirical covariance matrix of the bootstrap sample. Calculating 
\[ATS_{\vv}^{\vf,*}=N\cdot[\vC(\vf(\overline \vY^*)- \vf(\widehat \vv))]^\top[\vC(\vf(\overline \vY^*)- \vf(\widehat \vv))]/\tr(\vC\vJ_{\vf}(\overline \vY^*)\widehat\vSigma_{\dvech}^*  \vJ_{\vf}(\overline \vY^*)^\top\vC^\top) \]
repeatedly for sufficiently many independent realizations allows us to calculate their empirical $(1-\alpha)$ bootstrap quantile $q_{1-\alpha}^{\vh*}$.
Based on this, the bootstrap test is conducted, which controls the type-I error probability $\alpha$ asymptotically.\\\\

The parametric bootstrap from \cite{sattler2022} can also be adapted for this transformation in another way. To see this, only the covariance matrix $\vSigma$ is replaced by $\vSigma_{\dvech,\vf(\vv)}:=\vJ_{\vf}(\vv)\vSigma _{\dvech} \vJ_{\vf}(\vv)^\top$ and also the estimator $\widehat \vSigma$ by $\widehat \vSigma_{\dvech,\vf(\widehat \vv)}:=\vJ_{\vf}(\widehat\vv)\widehat\vSigma_{\dvech}  \vJ_{\vf}(\widehat\vv)^\top$. For a better understanding, we will recapitulate the procedure. First we generate a parametric bootstrap sample $\vY_{\vf, 1}^\dagger,...,\vY_{\vf, N}^\dagger\stackrel{i.i.d.}{\sim} \mathcal N_{p}\big(\vnull_p,\widehat \vSigma_{\dvech,\vf(\widehat \vv)}\big),$ for given realizations $\vX_{ 1},..., \vX_{ N}$, with estimators   $\widehat \vSigma_{\dvech}$ and $\widehat \vv$. From this bootstrap sample, we calculate the mean $\overline \vY_{\vf}^\dagger$ and the empirical covariance matrix $\widehat \vSigma_{\dvech,\vf}^\dagger$. The asymptotic normality follows as in \cite{sattler2022}, just with a different covariance matrix. This allows us to define the bootstrap version of the ATS by $ATS_{\vv}^{\vf,\dagger}=N[ \vC\overline \vY_{\vf}^\dagger  ]^\top[ \vC\ \overline \vY_{\vf}^\dagger  ]\big/\tr(\vC \widehat \vSigma_{\dvech,\vf}^\dagger \vC^\top)$. Based on a large number of repetitions, the corresponding conditional quantile and the test are obtained as before. The asymptotic correctness of this approach follows directly from the proof of Theorem 3 and Corollary 1 from \cite{sattler2022} by only replacing the covariance matrices.\\

It is worth mentioning that from a computational perspective, it would be more efficient to generate observations with the covariance matrix $\vC\widehat\vSigma_{\dvech,\vf(\widehat \vv)}\vC^\top$ and then avoid multiplications with $\vC$ in the computation of the bootstrap ATS.\\

Both bootstrap approaches can also be used identically for the vectorized correlation matrix by replacing $\vSigma_{\dvech}$ through $\vUpsilon_{\dvech}$.

\section{Proofs}
\begin{proof}[Proof of \Cref{TheoremFormulation}] 
Since the operation $\dvech(\cdot)$ on symmetric matrices is a bijection, we now switch to the equivalent definition
\[\dvech(\mathcal{V})=\{\vv(\vtheta)\in\R^{p}: \vv(\vtheta)= \vv_0+\theta_1 \vv_1+...+\theta_q \vv_q,\quad \vtheta = ( \theta_1, \dots, \theta_q)^\top\in\R^q\}\cap \dvech(\mathcal{COV}_{d\times d})\]
  with $\vv_0=\dvech(\vV_0),...,\vv_q=\dvech(\vV_q)$.
This is subset of an affine subspace defined as the linear hull of $\{\vv_1,\vv_2,...,\vv_q\}$ shifted by $\vv_0$, where we also know that this subspace has dimension $q$.
Now it is a well-known fact, see for example \cite{boyd2004}, Section~2.1, that, for each affine subspace, there exists a linear equation system which has this subspace as the solution set.\\
  As a repetition, we will sketch how $\vzeta$ and $\vC$ can be constructed for given $\vv_0,...,\vv_q$. In doing so, we follow the usual steps, which directly follows from the proof of the statement, but focus on standard basis vectors to simplify the implementation in usual programming languages. Each other basis could be used as well, together with a corresponding transition matrix.\\ \\
Since the dimension of $\mathcal{V}$ is only $q<p$, there exist standard basis vectors $\vb_1,...,\vb_{p-q}$ which, in combination with $\vv_1,...,\vv_q$, form a basis of $\R^p$. This allows us to define the matrix $\vB=(\vv_1,...,\vv_q,\vb_1,...,\vb_{p-q})^{-1}$. Now with the $j$-th standard basis vector $\ve_{j,p} \in \mathbb{R}^p$, we define $\vE\in\R^{(p-q)\times p}$ through $\vE=(\ve_{q+1,p},....,\ve_{p-q,p})$.
Finally, we get $\vC=\vE\cdot \vB$ and $\vzeta=\vC\cdot\vv_0$.\\\\
This equality between the affine subspace and the linear equation system, also holds if we consider only vectors within $\dvech(\mathcal{COV}_{d\times d})$, leading to the result for the subset $\mathcal{V}$.
\end{proof}

\begin{proof}[Proof of $\vV_0=\vnull_{d\times d} \Leftrightarrow \vzeta=\vnull_m$]
 From $\vzeta=\vC\vv_0$ it is clear that from $\vv_0=\vnull_m$ it  follows that $\vzeta=\vnull_m$. Therefore, we now consider the case of $\vzeta=\vnull_m$, which means that $\vnull_p$ solves the linear equation system. Since $\vnull_p$ is also in $\dvech(\mathcal{COV}_{d\times d})$ we knew $\vnull_p\in \dvech(\mathcal{V})$ and therefore there exist $\theta_1,...\theta_q$ with  $\vv_0+\theta_1\vv_1+...+\theta_q\vv_q=\vnull_p$. If now $\vv_0\neq \vnull_p$ this would mean that, with $(1,\theta_1,....\theta_q)$, we have a non-trivial linear combination which leads to $\vnull_p$, which contradicts the linear independence of $\vv_0,...,\vv_q$ and equivalently of $\vV_0,...,\vV_q$. For this reason, we conclude that $\vv_0=\vnull_p$, which completes the proof. 
\end{proof}

\begin{proof}[Proof of \Cref{Kof}]
The fact that $\vf$ is a differentiable function allows the use of the multivariate delta method (see, e.g. \cite{serfling1980}). Hence, it follows from \Cref{TheoremAlt} that
\[ \sqrt{N}\left(\vf(\widehat\vv)-\vf( \vv)\right)\stackrel{\mathcal D}{\longrightarrow}\mathcal{N}_{p}(\vnull_p,\vJ_{\vf}(\vv)\vSigma_{\dvech}  \vJ_{\vf}(\vv)^\top), \quad \text{as } N \to\infty.\]
Therefore, under the null hypothesis, we have
\[ \sqrt{N}\left(\vC\vf(\widehat\vv)-\vzeta\right)\stackrel{\mathcal D}{\longrightarrow}\mathcal{N}_{p}(\vnull_p,\vC\vJ_{\vf}(\vv)\vSigma_{\dvech}  \vJ_{\vf}(\vv)^\top\vC^\top), \quad \text{as } N \to\infty.\]
Since $\widehat \vSigma_{\dvech}$ is a consistent estimator for $\vSigma_{\dvech}$, it follows from the Continuous Mapping Theorem that $\tr(\vC\vJ_{\vf}(\widehat\vv)\widehat\vSigma_{\dvech}  \vJ_{\vf}(\widehat \vv)^\top\vC^\top)$ is a consistent estimator for $\tr(\vC\vJ_{\vf}(\vv)\vSigma_{\dvech}  \vJ_{\vf}(\vv)^\top\vC^\top)$.
In conclusion, Slutsky's theorem implies the following Central Limit Theorem under the null hypothesis:
$$ATS_{\vv}^{\vf}(\widehat{\vSigma}_{\dvech})\stackrel{\mathcal{D}}{\to}\sum_{k=1}^m \widetilde \lambda_k B_k$$
as $N\to\infty$, with $B_k\stackrel{i.i.d.}{\sim}\chi_1^2$ and $\widetilde \lambda_1,...,\widetilde \lambda_m$ are the eigenvalues of $(\vC\vJ_{\vf}(\vv)\vSigma_{\dvech}  \vJ_{\vf}(\vv)^\top\vC^\top)$.\\
This of course allows to use the Monte-Carlo approach to estimating quantiles, as it was sketched above.\\\\
As following from \Cref{Bootstraps}
$\sqrt{N}\vC (\vf(\overline \vY^*)- \vf(\widehat \vv))$ and $\sqrt{N} \vC\overline \vY_{\vf}^\dagger$ have given the data the same asymptotic distribution as $\sqrt{N}\left(\vC\vf(\widehat\vv)-\vzeta\right)$ under the null hypothesis. So because of the fact that the respective trace estimators are consistent, with Slutsky's theorem and the continuous mapping theorem, it follows that all three ATS have the same asymptotic distribution. Therefore, based on the bootstrap quantiles or the Monte-Carlo quantile, this leads to an asymptotically correct level $\alpha$ test.
\end{proof}

\begin{proof}[Proof of \Cref{TheoremFormulationextended}]

The basic idea of the proof was already sketched in \Cref{re:ftilde}, but the proof will now be exercised in more detail.
From \Cref{TheoremFormulation} we know the existence of $\vC$ and $\vzeta$ to represent the hypothesis of having the structure from  model~\eqref{eq:lcsmodel} by $\mathcal{H}_0^{\vv}:\vC\vv=\vzeta.$ This is now complemented with the restrictions of $\vTheta$ formulated through $\vf(\vv)=\vnull_\ell$. So to fulfill both at the same time, in accordance to
possible choice for $\widetilde \vC$, $\widetilde \vzeta$ and $\widetilde \vf$ would be \[\widetilde\vf(\vv)=\begin{pmatrix}\vv\\
 \vf(\vv)\end{pmatrix},\ \widetilde\vC=\begin{pmatrix}
\vC\\
\vI_{\ell}\end{pmatrix}\text{ and } \tilde \vzeta=\begin{pmatrix}\vzeta\\\vnull_\ell\end{pmatrix},\]
if $\vf$ fulfills the requirements. Then also $\widetilde\vf$ fulfills the requirements of  \Cref{Kof}.
This allows to formulate the corresponding hypothesis through 
$\mathcal{H}_0^{\vv}:\widetilde\vC\widetilde \vf(\vv)=\widetilde\vzeta$ 
and the dimension $m=(p-q)+\ell$ follows from the construction.
\end{proof}
\begin{proof}[Proof of \Cref{SameforCorrelation}]
We know that \[\sqrt{N}(\widehat \vr-\vr)\stackrel{\mathcal D}{ \longrightarrow}\vZ\sim\mathcal{N}_{p_u}\large(\vnull_{p_u},\vUpsilon_{\dvech}\large)  \quad \text{as \ } N \to \infty.\]

Another application of the delta method shows that, under the null hypothesis,
\[\sqrt{N}( \vf(\widehat \vr)-  \vf( \vr))\stackrel{\mathcal D}{\longrightarrow}\mathcal{N}_{p_u}(\vnull_{p_u},\vJ_{ \vf}(\vv)\widetilde \vA\vUpsilon  \widetilde \vA^\top\vJ_{  \vf}(\vv)^\top)\]
where $\vf:\R^{p_u}\to \R^{m}$ meets the required differentiability assumption. Under the null hypothesis $\vC\vr=\vzeta$, it therefore holds that
\[\sqrt{N}(\vC \vf(\widehat \vr)-\vzeta)\stackrel{\mathcal D}{\longrightarrow}\mathcal{N}_{p_u}(\vnull_{p_u},\vC\vJ_{  \vf}(\vv)\widetilde \vA\vUpsilon  \widetilde \vA^\top\vJ_{\vf}(\vv)^\top\vC^\top),\]

as $N \to \infty$.
With $\widehat \vUpsilon$ as a consistent estimator for $\vUpsilon$ from \cite{sattler2022} 
the results for 
$$ATS_{\vr}^{ \vf}:={N}(\vC\vf(\widehat \vr)-\vzeta)^\top(\vC\vf(\widehat \vr)-\vzeta)/\tr(\vC\vJ_{  \vf}(\widehat \vv)\widetilde \vA \widehat\vUpsilon  \widetilde \vA^\top\vJ_{  \vf}(\widehat \vv)^\top\vC^\top)$$
follow similarly as in the proof of \Cref{Kof}.
In analogy to \Cref{Bootstraps},  bootstrap approaches and a Monte-Carlo approach can be used to obtain appropriate quantiles.
\\\\
The result from \Cref{TheoremFormulation} holds identically for model \eqref{eq:lcrsmodel} by replacing $\dvech$ through $\dvech^-$ in the proof. Finally, the expansion from \Cref{TheoremFormulation} to \Cref{TheoremFormulationextended} follows with the same construction as for covariances.
\end{proof}
\subsection{Autoregressive covariance structure}

\subsubsection{Autoregressive covariance structure with function $\vg$}\label{2. Approach autoregressive}

The function $\vg:\R^{p}\to \R^{d-1}, $
is continuous and differentiable in $\vv$ with Jacobian matrix

\[
\vJ_{\vg}(\vx)=\sum\limits_{\ell=1}^{d-1} \ve_{{\ell},d-1} \tfrac{(d-\ell+1)}{(d-\ell)} 
\times \left[\sum\limits_{k=0}^{d-\ell-1} \ve_{a_{\ell+1}+k,p}^\top \cdot \frac{1}{\sum_{j=0}^{d-\ell} x_{a_{\ell}+j}}-\sum\limits_{k=0}^{d-\ell} \ve_{a_{\ell}+k,p}^\top\frac{\sum_{j=0}^{d-\ell-1} x_{a_{\ell+1}+j}}{ (\sum_{j=0}^{d-\ell} x_{a_{\ell}+j})^2}\right].
\]
Together with the formulation of the null hypothesis through
\[\mathcal{H}_0(AR): \vC_{AR}\ \begin{pmatrix}
 \vv\\
\vg( \vv)\end{pmatrix} =\begin{pmatrix}\veins_d\\\vnull_{p-1}
\end{pmatrix}.\]
this is all we need for using \Cref{Kof}.

\subsubsection{Autoregressive covariance structure with function $\vh$}\label{1. Approach autoregressive}

 To prepare a test for the autoregressive structure,  we first define the function
$$\vh:\R^{p}\to \R^{p}, \quad (x_1,...,x_p)\mapsto \left(x_1,...,x_d,|x_{d+1}|^{1/1},...,|x_{2d-1}|^{1/1},|x_{2d}|^{1/2},...,|x_{p}|^{1/(p-1)}\right).$$ 

When applied to a vectorized covariance matrix, this means that different roots for different secondary diagonals are used. Here, increasing roots are used (square root for the first secondary diagonal, cubic root for the second secondary diagonal, \emph{etc}), all applied to the absolute value. 
Under the null hypothesis, that is, for a vector $\vv$ which results from an autoregressive covariance matrix,  this leads to
\[
\vh(\vv)=(\underbrace{1,...,1}_{d \text{ times}}, \overbrace{|\rho|,...,|\rho|}^{(p-d) \text{ times}})^\top .\]
Under the null hypothesis of an autoregressive covariance matrix, every possible vectorized covariance matrix,  $\vv$, contains only positive entries or only negative entries, and therefore the function is continuous and differentiable on $\dvech(\mathcal{V})$.
Therefore, we calculate $\vJ_{\vh}$, the Jacobian matrix of the function $\vh$ as
\[\vJ_{\vh}(\vx)=\vI_{d}\oplus \left(\bigoplus\limits_{k=2}^d \diag(\eta(k,x_{a_{k}}),...,\eta(k,x_{a{_k}+d-k}))\right)\]
with

$\eta(k,x)= \tfrac{1}{k-1}\cdot x\cdot |x|^{\frac{3-2k}{k-1}}.$

 With a hypothesis matrix $\vC_{AR}=\vI_d\oplus \vC_{T}\oplus\vP_{p_u}$, we can formulate the  null hypothesis: 
\[\mathcal{H}_0^{\vv}:\vC_{AR}\begin{pmatrix}
    \vv\\
    \vh(\vv)
\end{pmatrix}=\begin{pmatrix}\veins_d\\\vnull_{p+p_u}\end{pmatrix},\]
which can be tested through an application of \Cref{Kof}.\\
Of course, through using the absolute values, we renounce information contained in the observations' signs. In doing so, we, in fact, check for a slightly larger structure, since, for example, negative values on the 2-th secondary diagonal are possible. This cannot be the case in autoregressive models. This potentially decreases the power, but it could be mended, for example, by incorporating signs into the function $\vh$. 

\subsubsection{First-order autoregressive}

Finally, the function $\vg$  allows us to test for one more version of the autoregressive structure called 'first-order autoregressive'. Here, the elements are given through $(\vV)_{ij}=\tau\cdot\rho^{|i-j|}$, with $\tau,\rho>0$. This is a more general case since the diagonal elements must be equal, but they are allowed to have values other than one. When we replace the $\vI_d$ in $\vC_{AR}$  by $\vP_d$ we change the hypothesis in a way to test for the equality of all diagonal elements. So we get the appropriate hypothesis matrix  $\vC_{FOAR}=\vC_T\oplus \vP_{d-2}$. Since the value $\tau$ is canceled out in the quotient, under the null hypothesis of a first-order autoregressive structure, it holds that  $\vC_{FOAR}( \vv^\top, \vg(\vv)^\top)^\top=\vnull_{p+d-2}$.

\subsection{Autoregressive correlation structure}\label{Autoregressive correlation structure}

Let 
\[\check\vg:\R^{p_u}\to \R^{d-2}, (x_1,...,x_{p_u})\mapsto \left(
\frac{ \tfrac 1{d-2}\cdot\sum_{k=0}^{d-3} x_{a_3+k-d}}{\tfrac 1 {d-1}\cdot\sum_{k=0}^{d-2} x_{a_2+k-d}},...,
\frac{\tfrac 1 1\cdot\sum_{k=0}^{0} x_{a_{d}+k-d}}{\tfrac 1 2\Cdot \sum_{k=0}^{1} x_{a_{d-1}+k-d}}\right),\]
consist of those components of $\vg$ which are relevant for testing 
the autoregressive structure of the correlation matrix and therefore not concern the diagonal elements.
For this version of $\vg$, the Jacobian matrix is  given through 
\[
\vJ_{\check\vg}(\vx)=\sum\limits_{\ell=2}^{d-1} \ve_{{\ell-1},d-2} \cdot \tfrac{(d-\ell+1)}{(d-\ell)} 
\times \left[\sum\limits_{k=\ell+1}^{d} \ve_{a_{\ell+1}-k,p_u}^\top \cdot \frac{1}{\sum_{j=\ell}^{d} x_{a_{\ell}-j}}-\sum\limits_{k=\ell}^{d} \ve_{a_{\ell}-k,p_u}^\top\frac{(\sum_{j=\ell+1}^{d} x_{a_{\ell+1}-j})}{ (\sum_{j=\ell}^{d} x_{a_{\ell}-j})^2}\right].
\]

In \cite{sattler2023correlation}, also a Taylor-based Monte-Carlo approach was presented, which showed good statistical properties in a simulation study.
With small adaptions regarding the other kind of vectorization, it can be used for this test with both functions $\widetilde \vg$, and $\widetilde \vh$ proceeding with the same steps as above.

\section{Additional simulations}\label{Additional simulations}
In addition to the simulations from the main part, we here additionally simulated $ATS_{\vv}^{\dagger,\vg}$  and $ATS_{\vv}^{\dagger,\vh}$ for all settings.  The results can be seen in \Cref{tab:autostrukturApp}.

           \begin{table}[htbp]
  \begin{small}
   \begin{center}
    \begin{tabular}{x{1.5pt}lx{1.5pt}c|c|c|cx{1.5pt}c|c|c|cx{1.5pt}c|c|c|cx{1.5pt}}\specialrule{1.5pt}{0pt}{0pt}
     \rowcolor{ashgrey}  
  &\multicolumn{4}{|cx{1.5pt}}{ATS-Para-$\vg^*$}&\multicolumn{4}{|cx{1.5pt}}{ATS-Para-$\vg^\dagger$}&\multicolumn{4}{cx{1.5pt}}{ATS-$\vg$}\\ \specialrule{1.5pt}{0pt}{0pt}   
\rowcolor{gainsboro}N&25&50&100&250&25&50&100&250&25&50&100&250
       \\\specialrule{1.5pt}{0pt}{0pt}
\rowcolor{lightyellow}$t_9$ &3.98 & 3.86 & \bf{4.62} & \bf{4.95} & 1.50 & 1.78 & 2.79 & 3.75 & 1.76 & 1.89 & 2.61 & 3.67 \\     \hline
\rowcolor{gainsboro}Normal  &  3.80 & 3.94 & \bf{4.70} & \bf{4.97} & 1.39 & 1.77 & 2.53 & 3.66 & 1.67 & 1.87 & 2.49 & 3.81 \\  \hline
\rowcolor{lightyellow}  Skew normal &4.37 & 4.21 & 4.37 & \bf{5.00} & 1.51 & 1.86 & 2.57 & 3.59 & 1.79 & 2.03 & 2.51 & 3.62 \\  \hline
\rowcolor{gainsboro}Gamma &  \bf{5.15} & \bf{5.00} & \bf{4.66} & \bf{4.70} & 2.41 & 2.48 & 2.77 & 3.57 & 2.70 & 2.62 & 2.75 & 3.64 \\    \specialrule{1.5pt}{0pt}{0pt}
\multicolumn{9}{x{1.5pt}cx{1.5pt}}{}\\\specialrule{1.5pt}{0pt}{0pt}
     \rowcolor{ashgrey}     &\multicolumn{4}{|cx{1.5pt}}{ATS-Para-$\vh^*$}&\multicolumn{4}{|cx{1.5pt}}{ATS-Para-$\vh^\dagger$}&\multicolumn{4}{cx{1.5pt}}{ATS-$\vh$}\\\specialrule{1.5pt}{0pt}{0pt}  
\rowcolor{gainsboro}N&25&50&100&250&25&50&100&250&25&50&100&250
       \\\specialrule{1.5pt}{0pt}{0pt}
\rowcolor{lightyellow}$t_9$ &5.50 & \bf{5.33} & \bf{4.96} & \bf{4.93} & 3.23 & 3.01 & 3.99 & 4.53 & 3.50 & 3.29 & 3.99 & \bf{4.63} \\   \hline
\rowcolor{gainsboro}Normal   &5.61 & 5.82 & 5.54 & 5.58 & 2.98 & 3.37 & 4.17 & \bf{4.77} & 3.27 & 3.57 & 4.18 & \bf{4.87} \\   \hline
 \rowcolor{lightyellow} Skew normal &6.49 & 5.92 & 5.47 & \bf{5.41} & 3.54 & 3.75 & 4.15 & \bf{4.82} & 3.93 & 3.91 & 4.13 & \bf{4.90} \\   \hline
\rowcolor{gainsboro}Gamma &  6.75 & 5.69 & \bf{4.83} & \bf{4.65} & 4.12 & 3.68 & 3.72 & 4.16 & \bf{4.68} & 3.87 & 3.79 & 3.99 \\   \specialrule{1.5pt}{0pt}{0pt}

\end{tabular}
\end{center}
  \caption{Simulated type-I-error rates ($\alpha=5\%$) in percentage, for testing whether the covariance matrix has an autoregressive structure, with ATS based on two kinds of parametric bootstraps and a Monte-Carlo approach. The observation vectors have dimension 5, covariance matrix $(\vV_1)_{ij}=0.65^{|i-j|}$ and different distributions and sample sizes are considered.}
    \label{tab:autostrukturApp}
\end{small}
    \end{table}

It can be seen that $ATS_{\vv}^{\dagger,\vg}$ continually perform worse than $ATS_{\vv}^{*,\vg}$ and even the corresponding Monte-Carlo versions; identically for the function $\vh$.
Therefore, regarding the type-I error rate of $ATS_{\vv}^{\dagger,\vg}$ and $ATS_{\vv}^{\dagger,\vh}$, it makes less sense to use this approach here. \\

Furthermore, we compared the power of all three tests based on the function $\vg$ and $\vh$ for the normal and the $t_9$ distribution. Comparing the results as displayed in \Cref{tab:PAR2} and \Cref{tab:PAR3}, it can be seen that the power is slightly better under the normal distribution than for the $t_9$ distribution. 
Moreover, the power of $\vg^*$  is about five percentage points higher than for $\vg^\dagger$  while the difference is smaller for  $\vh^*$   and $\vh^\dagger$. The most remarkable difference in power is between the approaches based on the functions $\vg$ and $\vh$. Despite its less liberal behaviour, the approach based on the function $\vg$ leads to a significantly higher power, often more than doubling the power of the tests based on the function $\vh$.

   \begin{table}[htbp]
    
 \begin{center}
    \begin{tabular}{x{1.5pt}lx{1.5pt}r|r|r|r|r|r|r|r|r|r|cx{1.5pt}}\specialrule{1.5pt}{0pt}{0pt}     
\rowcolor{ashgrey}  \hspace{0.7cm}$\delta$&0.0&0.1&0.2&0.3&0.4&0.5&0.6&0.7&0.8&0.9&1.0
       \\\specialrule{1.5pt}{0pt}{0pt}
       \rowcolor{gainsboro}ATS-Para-$\vg^*$  & 4.8 & 6.9 & 12.4 & 24.2 & 40.1 & 64.0 & 80.6 & 91.8 & 97.5 & 99.6 & 100.0 \\  \hline 
\rowcolor{lightyellow}ATS-Para-$\vg^\dagger$ & 3.8 & 4.9 & 9.7 & 19.4 & 34.2 & 56.4 & 76.1 & 89.1 & 96.9 & 99.6 & 99.8 \\    \hline
\rowcolor{gainsboro} ATS-$\vg$  & 3.5 & 5.1 & 10.0 & 19.6 & 35.0 & 57.3 & 76.2 & 89.5 & 96.8 & 99.6 & 99.8 \\  \specialrule{1.5pt}{0pt}{0pt}

\rowcolor{lightyellow}ATS-Para-$\vh^*$ & 5.4 & 5.6 & 7.4 & 9.3 & 13.7 & 21.0 & 30.0 & 39.1 & 50.2 & 61.2 & 67.1 \\   \hline 
\rowcolor{gainsboro}ATS-Para-$\vh^\dagger$& 4.5 & 5.1 & 6.9 & 8.6 & 12.6 & 18.5 & 28.7 & 35.7 & 45.8 & 56.8 & 63.3 \\    \hline
\rowcolor{lightyellow} ATS-$\vh$ & 4.2 & 4.8 & 6.9 & 8.5 & 12.5 & 18.1 & 27.7 & 34.2 & 45.1 & 55.7 & 61.3 \\  \specialrule{1.5pt}{0pt}{0pt}

\end{tabular}
  \caption{Simulated power ($\alpha=5\%$) in percentage for testing whether the covariance matrix has an autoregressive structure, with ATS based on parametric bootstraps and Monte-Carlo simulation. The observation vectors have dimension 5, covariance matrix $\vV_\delta=(1-\delta) \vV_1+ \delta\vV_2$ and normal distribution.}
    \label{tab:PAR2}
    \end{center}

     \end{table}
   \begin{table}[htbp]
    
 \begin{center}
    \begin{tabular}{x{1.5pt}lx{1.5pt}r|r|r|r|r|r|r|r|r|r|cx{1.5pt}}\specialrule{1.5pt}{0pt}{0pt}    
\rowcolor{ashgrey}  \hspace{0.7cm}$\delta$&0.0&0.1&0.2&0.3&0.4&0.5&0.6&0.7&0.8&0.9&1.0
       \\\specialrule{1.5pt}{0pt}{0pt}
       \rowcolor{gainsboro}ATS-Para-$\vg^*$ &  5.1 & 6.6 & 10.3 & 19.0 & 30.4 & 49.7 & 65.5 & 83.5 & 93.0 & 97.4 & 98.9 \\    \hline
\rowcolor{lightyellow}ATS-Para-$\vg^\dagger$  &3.7 & 4.9 & 7.7 & 13.7 & 25.8 & 42.5 & 59.5 & 79.3 & 91.0 & 96.0 & 98.6 \\    \hline
\rowcolor{gainsboro} ATS-$\vg$  &3.8 & 4.8 & 8.0 & 14.2 & 25.7 & 43.5 & 59.1 & 80.7 & 90.8 & 96.7 & 98.6 \\  \specialrule{1.5pt}{0pt}{0pt}
\rowcolor{lightyellow}ATS-Para-$\vh^*$ &5.2 & 5.9 & 6.6 & 6.5 & 8.6 & 13.1 & 19.4 & 31.0 & 38.2 & 45.9 & 56.8 \\   \hline
\rowcolor{gainsboro}ATS-Para-$\vh^\dagger$  & 5.0 & 5.6 & 6.1 & 6.0 & 7.1 & 11.8 & 17.2 & 27.9 & 34.0 & 41.6 & 51.7 \\   \hline
\rowcolor{lightyellow} ATS-$\vh$  &4.7 & 5.4 & 5.8 & 5.6 & 7.0 & 11.5 & 17.3 & 27.1 & 32.8 & 40.1 & 50.8 \\  \specialrule{1.5pt}{0pt}{0pt}

\end{tabular}
  \caption{Simulated power ($\alpha=5\%$) in percentage for testing whether the covariance matrix has an autoregressive structure, with ATS based on parametric bootstraps and Monte-Carlo simulation. The observation vectors have dimension 5, covariance matrix $\vV_\delta=(1-\delta) \vV_1+ \delta\vV_2$ and $t_9$ distribution.}
    \label{tab:PAR3}
    \end{center}
     \end{table}

\bibliographystyle{apalike}
\bibliographystyle{unsrtnat}
\newpage

\addcontentsline{toc}{chapter}{Bibliography}
\bibliography{Literatur}

\end{document}